\documentclass[10pt,pre,twocolumn,amsmath,amssymb,superscriptaddress]{revtex4-1}

\usepackage{xr}

\usepackage{url}
\usepackage{hyperref}
\hypersetup{
    colorlinks,
    citecolor=blue,
    filecolor=blue,
    linkcolor=blue,
    urlcolor=blue
}

\usepackage{tabularx}
\usepackage{makecell}

\usepackage{mathrsfs,ragged2e}
\usepackage{verbatim}
\usepackage{graphicx}	
\usepackage{dcolumn}	
\usepackage{bm}		
\usepackage{bbm,upgreek}
\usepackage{mathtools}
\usepackage{physics}
\usepackage{dsfont}
\usepackage{cancel}
\usepackage{xcolor}
\usepackage{harpoon}
\usepackage{accents}
\usepackage[normalem]{ulem}
\usepackage{mathtools,amscd}

\usepackage{color}
\usepackage[cal=boondoxo]{mathalfa}
\usepackage{palatino}

\makeatletter
\def\maketitle{
\@author@finish
\title@column\titleblock@produce
\suppressfloats[t]}
\makeatother

\usepackage{graphicx}

\usepackage{dsfont}
\usepackage{cancel}
\usepackage{xcolor}
\usepackage{harpoon}
\usepackage{accents}
\usepackage{colortbl}
\usepackage{tablefootnote}

\usepackage{ulem} 

\newcommand{\AB}{\textit{A.~baylyi}}

\newcommand*{\idea}[1]{\medskip \noindent \textbf{#1}}

\newcommand*{\iidea}[1]{\medskip \noindent \textit{#1}}

\DeclareRobustCommand{\SkipTocEntry}[5]{}
\newcommand{\TocSkip}{\addtocontents{toc}{\SkipTocEntry}}

\newcommand{\beginsupplement}{%
        \setcounter{table}{0}
        \renewcommand{\thetable}{S\arabic{table}}%
        \setcounter{figure}{0}
        \renewcommand{\thefigure}{S\arabic{figure}}%
        \renewcommand{\theHfigure}{S\arabic{figure}}
        \setcounter{equation}{0}
        \renewcommand{\theequation}{S\arabic{equation}}
        \renewcommand{\theHequation}{S\arabic{equation}}
        \renewcommand\thesection{\arabic{section}}
     }

\begin{document}


\title{Protein overabundance is driven by growth robustness}

\author{H. James Choi}
\author{Teresa W.\ Lo}
\author{Kevin J.\ Cutler}
\author{Dean Huang}
\affiliation{Department of Physics, University of Washington, Seattle, Washington 98195, USA}
\author{W.\ Ryan Will}
\affiliation{Department of Laboratory Medicine and Pathology, University of Washington, Seattle, Washington 98195, USA}
\author{Paul A.\ Wiggins}
\email{pwiggins@uw.edu}
\affiliation{Department of Physics, University of Washington, Seattle, Washington 98195, USA}
\affiliation{Department of Microbiology, University of Washington, Seattle, Washington 98195, USA}
\affiliation{Department of Bioengineering, University of Washington, Seattle, Washington 98195, USA}

\begin{abstract}

Protein expression levels optimize cell fitness: Too low an expression level of essential proteins will slow growth by compromising essential processes; whereas overexpression slows growth by increasing the metabolic load. This trade-off na\"ively predicts that cells maximize their fitness by sufficiency, expressing just enough of each essential protein for function. We test this prediction in the naturally-competent bacterium \textit{Acinetobacter baylyi} by characterizing the proliferation dynamics of essential-gene knockouts at a single-cell scale (by imaging) as well as at a genome-wide scale (by TFNseq).  In these experiments, cells proliferate for multiple generations as target protein levels are diluted from their endogenous levels. This approach facilitates a proteome-scale analysis of protein overabundance. 
As predicted by the Robustness-Load Trade-Off (RLTO) model, we find that roughly 70\% of essential proteins are overabundant and that  overabundance  increases as the expression level decreases, the signature prediction of the model. These results reveal that robustness plays a fundamental role in determining the expression levels of essential genes and that overabundance is a key mechanism for ensuring robust growth.

\end{abstract}



\maketitle


Understanding the rationale for protein expression levels is a fundamental question in  biology with broad implications for understanding cellular function \cite{Albe_2002_book}. Measured expression levels appear to be paradoxically both \textit{optimal} and \textit{overabundant}. For instance, repeated investigations support the idea that gene expression levels optimize cell fitness \cite{Dekel:2005fu,Lalanne:2021xs}. Since the overall metabolic cost of protein expression is large  \cite{Lengeler1998,Kafri:2016fu}, fitness optimization would seem to imply that protein levels should satisfy a \textit{Goldilocks condition}: Expression levels should be \textit{just high enough} to achieve the required protein activity \cite{Hausser:2019fi,Belliveau:2021pj}.  However, a range of approaches suggest that many essential genes are expressed in vast excess of the levels required for function \cite{Peters:2016jf,Donati:2021kq,Belliveau:2021pj}. 
How can expression levels be at once optimal with respect to fitness as well as in excess of what is required for function?

The cell faces a complex regulatory challenge: Even in a bacterium, there are between five and six hundred essential proteins, each of which is required for growth \cite{Baba:2008vn}. How does the cell ensure the robust expression of each essential factor? We recently argued that the stochasticity of gene expression processes fundamentally shape the principles of central dogma regulation, including the optimality of protein overabundance \cite{Lo2024}.  Specifically, we proposed a quantitative model, the Robustness-Load Trade-Off (RLTO) model, which makes a parameter-free prediction of protein overabundance as a function of gene transcription level \cite{Lo2024}. The optimality of overabundance can be understood as the result of a highly-asymmetric fitness landscape: the fitness cost of essential protein underabundance, which causes the arrest of essential processes, is far greater than the fitness cost of essential protein overabundance, which leads to slow growth by increasing the metabolic load. However, critical model assumptions and predictions remain untested which is the motivation for the current study. Here, we will quantitatively measure the fitness landscape with repect to protein abundance and determine the level of overabundance for all essential proteins in the bacterium \textit{Acinetobacter baylyi}.

\TocSkip\section*{Results}

\begin{figure*}
  \centering
    \includegraphics[width=0.98\textwidth]{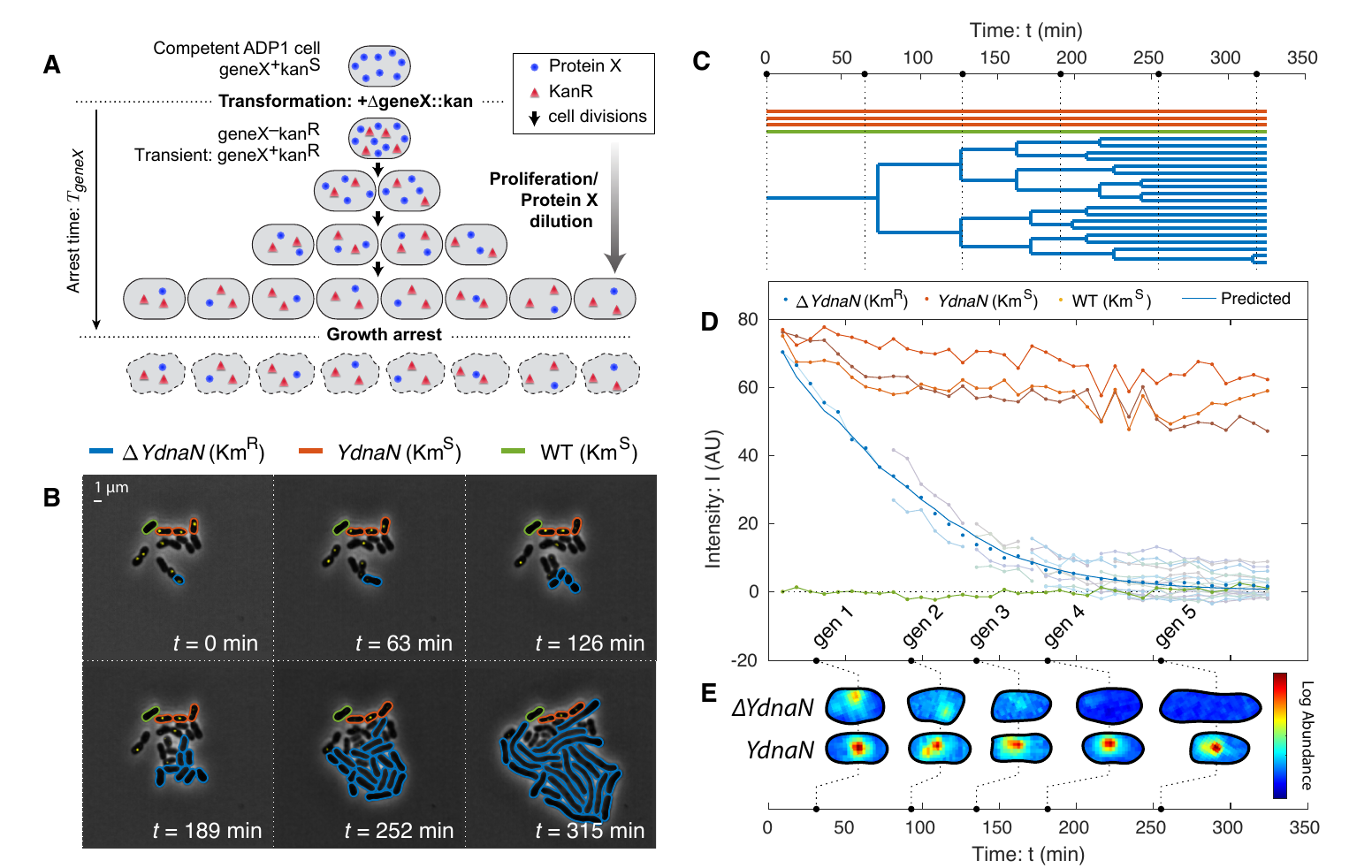}
      \caption{
      \textbf{Knockout-depletion experiments. Panel A: Experimental schematic.} Competent ADP1 cells are transformed with $\Delta$\textit{geneX::kan}. Untransformed cells arrest immediately on selective media. Transformed cells proliferate, but cease protein X expression (blue circles) while expressing Kan (red triangles). Existing protein X abundance is diluted as cells proliferate. For essential genes, cell growth continues until protein levels are diluted to the threshold level required for growth, after which growth arrests.    
      \label{fig:growthCont}  \textbf{Panel B \& C: Visualization of knockout depletion.}
      The  fluorescent fusion \textit{YPet-dnaN} to essential gene \textit{dnaN} is knocked outed at $t = 0$. Cell proliferation is visualized using phase-contrast microscopy while protein abundance is measured by fluorescence microscopy (yellow).  Transformed cells ($\Delta$\textit{YdnaN}, blue) have a Km$^R$ allele and can proliferate over several generations before arrest; however, untransformed cells (\textit{YdnaN}, orange) and wild-type cells (WT, green) were both kanamycin sensitive and therefore arrested immediately. 
      \textbf{Panel C: Lineage tree.}   
      Black dotted lines represent time points shown in Panel B.   \textbf{Panel D: Target protein is diluted by proliferation.} Protein concentration is measured by integrated fluorescence. Arrested \textit{YdnaN} cells maintain protein abundance, whereas proliferating transformed cells ($\Delta$\textit{YdnaN},  blue) show growth-induced protein depletion. The protein concentration over all transformed progeny (blue points) are consistent with the dilution-model prediction (solid blue).  \textbf{Panel E: Protein function is robust to dilution.}  Representative single-cell images of transformed ($\Delta$\textit{YdnaN}) and untransformed (\textit{YdnaN}) cells are shown for successive time points. The YPet-DnaN fusion shows punctate localization, consistent with function, even as protein abundance is depleted. No puncta are observed in the last generation and the cells form filaments, consistent with replication arrest. 
      }
      \label{fig:dilution}
\end{figure*}

\idea{Natural competence facilitates knockout-depletion.} To characterize the fitness landscape for essential gene expression, we must deplete the levels of essential proteins. Both degron- and CRISPRi-based approaches have been applied; however, these approaches require careful characterization of protein levels \cite{McGinness:2006gn,Davis:2011mf,Cameron:2014yt,Liu:2017tm,Peters:2016jf} and  introduce significant cell-to-cell variation on top of the endogenous noise which further obscures the underlying fitness landscape \cite{Franks:2024uh}. To circumvent these difficulties, we will use an alternative approach: \textit{knockout-depletion} in the naturally competent bacterium \textit{A.\ baylyi} ADP1 \cite{Bailey:2019tp,Gallagher:2020sp}. 
In this approach, cells are transformed with a \textit{geneX::kan} knockout cassettes targeting essential gene X, carrying a kanamycin resistance allele Km$^R$. (See Fig.~\ref{fig:growthCont}A.) Cells that are not transformed arrest immediately on selective media. The crux of the approach is that transformants remain transiently \textit{geneX}$^+$, due to the presence of already synthesized target protein X, even after the transcription of the target \textit{geneX} stops. Growth can continue, diluting protein X abundance, as long as this residual abundance remain sufficient for function. 
The success of the knockout-depletion approach is dependent on the extremely high transformation efficiency of \textit{A.~baylyi}.


\begin{figure*}
  \centering
    \includegraphics[width=0.98\textwidth]{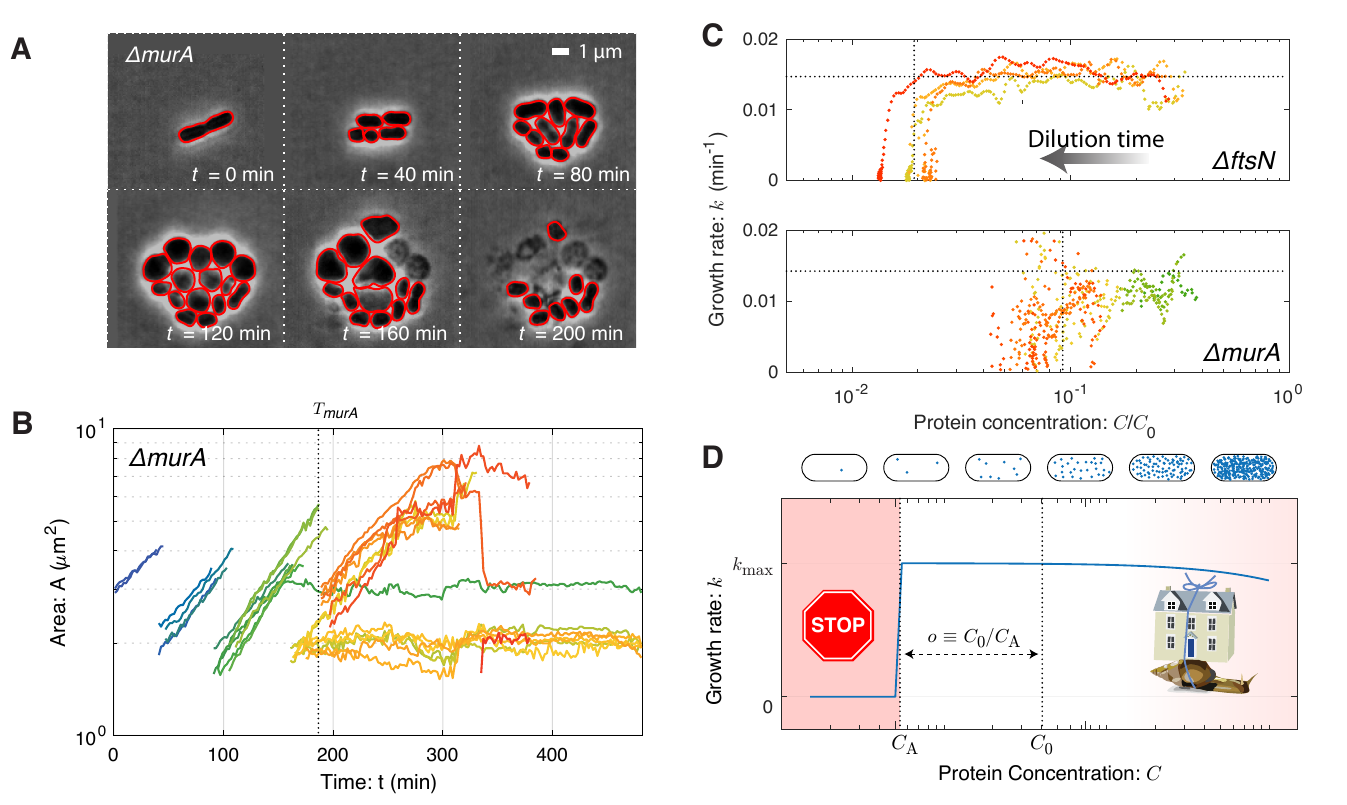}
      \caption{
     \textbf{The fitness landscape. Panel A: Visualization of growth in a \textit{murA} knockout.}  Essential gene \textit{murA} is knocked out at $t = 0$ and cell proliferation is visualized by phase-contrast microscopy.  Red outlines represent the Omnipose cell segmentation. Cell proliferation continues for multiple generations after deletion. 
     \textbf{Panel B: Quantitative analysis of cell proliferation with single-cell resolution.} Cell area (log scale) as a function of time  for the \textit{murA} deletion. The log-slope represents the single-cell growth rate.  The vertical dotted line represents the arrest time at which cell growth slows to cell arrest. 
     \textbf{Panel C: Growth rate as a function of protein depletion for $\Delta$\textit{ftsN} and $\Delta$\textit{murA}.} In both essential gene deletions, the growth rate is observed to obey the step-like-dependence, transitioning between wild-type growth to arrest at the vertical dotted lines. We define the critical dilution as $o\equiv C_0/C_A$ where $C_A$ is the protein concentration at arrest.
      \textbf{Panel D: The fitness landscape is threshold-like.} Motivated by single-cell growth data, cell fitness is modeled using the Robustness-Load Trade-Off model (RLTO). In the model, there is a metabolic cost of protein expression which favors low expression; however, growth arrests for protein concentration $C$ smaller than the threshold level $C_A$ (red).  The relative metabolic cost of overabundance is small relative to the cost of growth arrest due to the large number of proteins synthesized, resulting in a highly asymmetric fitness landscape \cite{Lo2024}.  
     } \label{fig:singlecell}
\end{figure*}

\idea{Target proteins are depleted by dilution.} A key untested assumption in the experimental design of the knockout-depletion approach is that target protein translation stops after transformation, and that the protein abundance is depleted by dilution.  The model predicts that the protein concentration is: 
\begin{equation}
C(t) = \textstyle C_0\cdot {V_0}/{V(t)}, \label{eqn:dil}
\end{equation} 
where $C_0$ and $V_0$ are the concentration and volume of the progenitor cell at deletion and $V(t)$ is the total volume of the progeny. 
To test the predicted protein depletion hypothesis, we designed a knockout-depletion experiment to target a protein we had previously studied that can be visualized using a fluorescent fusion and whose localization is activity dependent: the essential replication gene \textit{dnaN}, whose gene product is the $\beta$ sliding clamp \cite{Mangiameli2017b,Mangiameli2017,Mangiameli2018}. We  constructed a N-terminal fluorescent fusion to \textit{dnaN} using YPet in \textit{A.~baylyi} at the endogenous locus. The resulting mutant (YdnaN)
had no measurable growth defect under our experimental conditions.
We then knocked out the \textit{YPet-dnaN} fusion, yielding $\Delta$\textit{dnaN}, and characterized the protein levels by quantifying YPet-DnaN abundance by fluorescence. 
The experimentally measured fluorescence intensity is consistent with the dilution model (Eq.~\ref{eqn:dil}), as expected. (See Fig.~\ref{fig:dilution}D.)  We therefore conclude that knockout-depletion experiments are consistent with the experimental design shown schematically in Fig.~\ref{fig:growthCont}A.


\idea{Replication persists during DnaN depletion.} A key subhypothesis of the overabundance model for transient growth is that target protein function continues as the target protein abundance is depleted.
An alternative hypothesis for transient growth of the $\Delta$\textit{dnaN} strain is a high initial chromosomal copy-number that is partitioned between daughter cells, even after the replication process itself arrests due to target protein depletion \cite{Lengeler1998,Cooper:2000af}.
The imaging-based knockout-depletion experiment tests this hypothesis as well.  The localization of DnaN is dependent on activity:  During ongoing replication, DnaN is localized in puncta corresponding to replisomes, whereas in the absence of active replication,  DnaN has diffuse localization \cite{Lemon:1998pl,Reyes-Lamothe:2010nv,Mangiameli2017b,Mangiameli2017,Mangiameli2018}. During the knockout-depletion experiment, we observed YPet-DnaN puncta persist as the targeted fusion was depleted (Fig.~\ref{fig:dilution}DE), consistent with replication activity after dilution. Only after the YPet-DnaN puncta disappear do the cells begin to adopt the $\Delta$\textit{dnaN} phenotype: cell filamentation (Fig.~\ref{fig:dilution}BE).
We therefore conclude that function (replication) is robust to significant target protein (DnaN) dilution.

\begin{table*}[]
    \centering
\resizebox{1 \textwidth}{!}{    \begin{tabularx}{\textwidth}{ l|X|r|>{\columncolor{blue!10}}r|>{\columncolor{blue!10}}r>{\columncolor{blue!10}}r r}
  \multicolumn{3}{l|}{\ }                                  & \multicolumn{4}{l}{\bf Log Overabundance:}                                                                                 \\
  \multicolumn{2}{l|}{\ }     & \multicolumn{1}{l|}{Message number:}   & \multicolumn{1}{l}{\cellcolor{white}TFNseq}  & \multicolumn{2}{|l}{Imaging-based}                                &         \\
  \multicolumn{2}{l|}{\ }     &  \multicolumn{1}{l|}{$\mu_m$ (mRNA mol-}  & \multicolumn{1}{l}{\cellcolor{white}{Replication}}   & \multicolumn{1}{|l}{\cellcolor{white}{Elongation}} & \cellcolor{white}{Septation}       &       \\
{\bf Gene}       &   \multicolumn{1}{l|}{\bf Annotated gene function:} &  \multicolumn{1}{l|}{ecules/cell cycle)} &  \multicolumn{1}{l}{$\log_{10} o$}             & \multicolumn{1}{|l}{$\log_{10} o$} & \multicolumn{1}{l}{$\log_{10} o$}  & ($N_C$, $N_P$)  \\
   \hline  
   \hline
   \textit{dnaA} & Regulation of replication initiation       &  30  & 0.02$\pm$ 0.02  & 0.7 $\pm$ 0.1     &  0.0$\pm$ 0.2   & (4,4)    \\
   \textit{dnaN} & Replication beta sliding clamp             &  49  & 1.5 $\pm$ 0.1   & 2.0 $\pm$ 3.0     &  1.4$\pm$ 0.1   & (134,8)   \\
   \textit{ftsN} & Essential cell division/septation protein  &  20  & 2.6 $\pm$ 0.1   & 1.8 $\pm$ 0.2     &  0.6$\pm$ 0.2   & (19,5)   \\
   \textit{murA} & Cell wall precursor synthesis              &  26  & 0.7 $\pm$ 0.5   & 1.1 $\pm$ 0.1     &  0.9$\pm$ 0.2   & (16,4)   \\
   \hline
    \end{tabularx}}
    \caption{\textbf{Measured  overabundance  for sequencing- versus imaging-based approaches.} The overabundance was determined by both sequencing- and imaging-based approaches. For the imaging-based approach, we show two measurements based on different metrics for arrest: The first is based on the arrest of cell elongation, as defined by Eq.~\ref{eqn:grelong},  and the second is based on the arrest of the septation process, as visualized by microscopy. \label{tab:overab}}   
\end{table*}

\idea{Many essential knockouts undergo transient growth.} To understand the generic consequences of essential protein depletion, we used the imaging-based knockout-depletion experiments to explore essential genes with a range of functions.  
We initially targeted four essential genes: the replication initiation regulator gene \textit{dnaA} (\href{https://youtu.be/EjYRzKXhfe0}{movie}), the beta-clamp gene \textit{dnaN} (\href{https://youtu.be/i58BSBwx_hw}{movie}), the cell-wall-synthesis gene \textit{murA} (\href{https://youtu.be/E7f_Pi1inkw}{movie}), and septation-related gene \textit{ftsN} (\href{https://youtu.be/HFGs1jlVJ7c}{movie}), as well as a non-essential IS element with no phenotype as a negative control (\href{https://youtu.be/FgIF9LsCVmA}{movie}). (Representative frame mosaic images and cytometry appear in Supplementary Material Sec.~\ref{sec:smTFNseq}.) In each case, transformants continued to proliferate through multiple cell-cycle durations \cite{Bailey:2019tp} and are therefore consistent with the essential protein overabundance hypothesis. 
However, in Ref.~\cite{Bailey:2019tp}, we were unable to perform a quantitative single-cell analysis of these time-lapse experiments since existing segmentation packages failed to segment the observed morphologies \cite{Cutler2022}. We therefore developed the \textit{Omnipose} package,  which facilitated quantitative analysis of the growth dynamics with single-cell resolution \cite{Cutler2022}.  (See Fig.~\ref{fig:singlecell}A.)

\idea{The fitness landscape is threshold-like.}
A key input to the RLTO model is the fitness landscape (growth rate) as a function of protein abundance.
Omnipose segmentation facilitates the measurement of single-cell growth rates from the time-lapse imaging experiments. We focus first on the single-cell areal growth rate: 
\begin{equation}
k(t) = \textstyle\frac{\rm d}{{\rm d}t} \ln A(t), \label{eqn:growthrate}
\end{equation} 
where $A(t)$ is the area of the cell at time $t$. This areal growth rate is more convenient than a cell-length based rate since we avoid the necessity of defining cell length for unusual cell morphologies like those observed in the $\Delta$\textit{murA} mutant. Fig.~\ref{fig:singlecell}B shows representative knockout-depletion dynamics of cell area for the essential-gene target \textit{murA}. The log slope remains constant for multiple generations, consistent with a constant growth rate, even as the gene targeted is depleted over multiple cell cycles.  
By combining the dilution model (Eq.~\ref{eqn:dil})  and  the growth rate (Eq.~\ref{eqn:growthrate}), a single knockout-depletion measurement determines the growth rate for a range of protein abundances between wild-type abundance and those realized at growth arrest. This fitness landscape is shown for the MurA and FtsN proteins in Fig.~\ref{fig:singlecell}C. For all four mutants, the areal growth rate is roughly constant for multiple generations before undergoing a rapid transition to growth arrest. 

\idea{Protein overabundance.}
We will define the overabundance as the ratio of protein concentration in wild-type cells ($C_0$) to the concentration at cell arrest ($C_A$):
\begin{equation}o \equiv C_0/C_{\rm A}, \label{eqn:grelong}
\end{equation} 
as shown in Fig.~\ref{fig:singlecell}D. (Supplementary Material Sec.~\ref{sec:sm_imaging} gives a detailed description of the inferred overabundance from single-cell data.)
The measured overabundance for the four mutants imaged by microscopy is summarized in Tab.~\ref{tab:overab}, using three distinct metrics for growth. We conclude that for each gene, with the exception of \textit{dnaA}, rapid growth continues after the knockout due to the vast overabundance of the target protein.


\idea{The RLTO model predicts protein overabundance.} 
The RLTO model explicitly analyzes the trade-off between growth robustness to noise and metabolic load and predicts the optimal central-dogma regulatory principles \cite{Lo2024}. Critically, the model incorporates the observed threshold-like dependence of growth rate on protein abundance (Fig.~\ref{fig:singlecell}CD). 
The model quantitatively predicts protein overabundance with a signature feature: high-expression genes have low protein overabundance ($o \approx 1$) due to the high metabolic cost of increasing expression and low inherent noise of high expression genes; however, low-expression genes have high overabundance ($o \gg 1$) due to the low metabolic cost of increasing expression and the high inherent noise of low expression genes. (See Supplemental Material Sec.~\ref{sec:RLTO} for a more detailed description of the model.) 

\begin{figure*}
  \centering
    \includegraphics[width=1\textwidth]{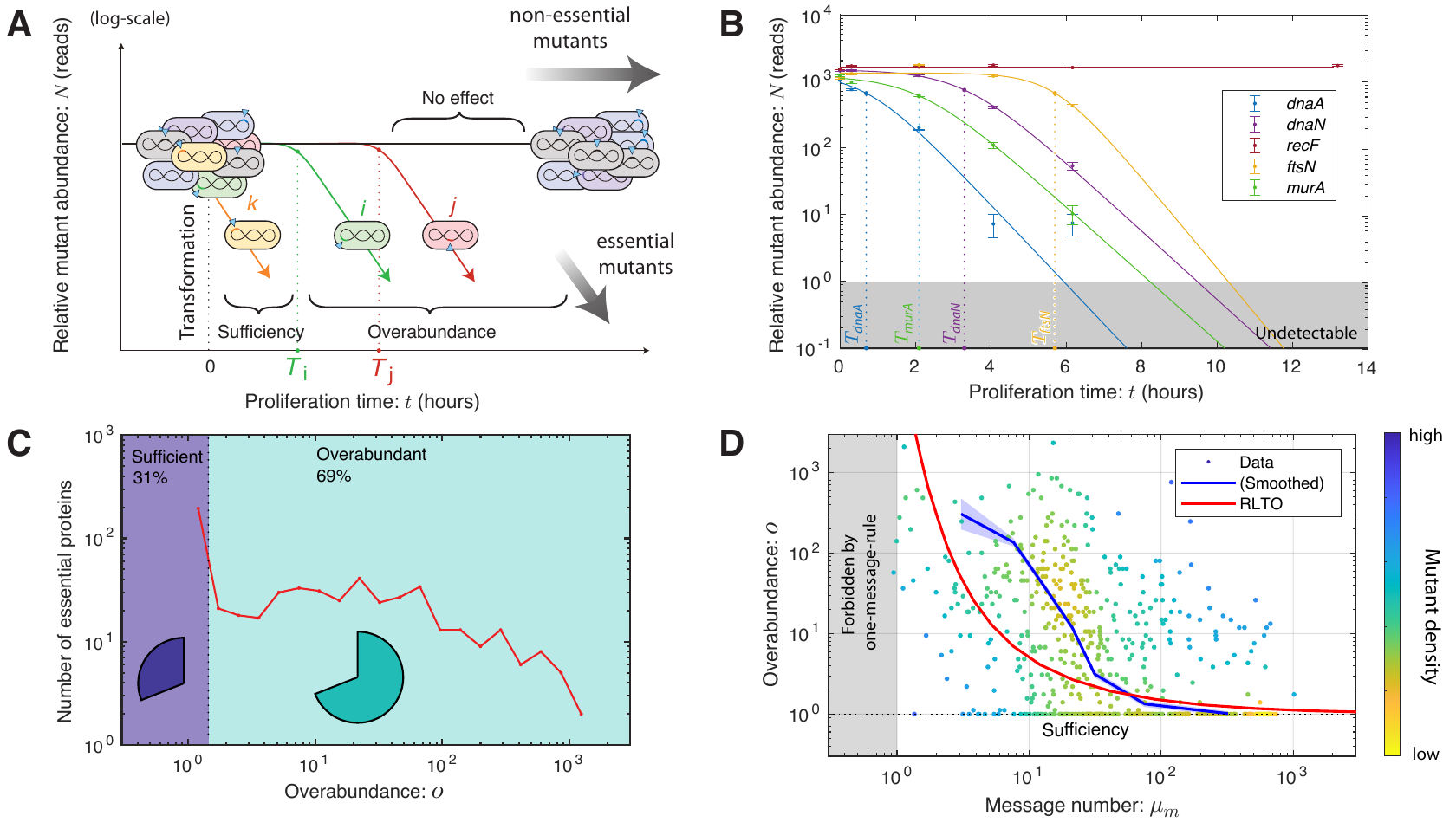}
      \caption{ \textbf{A proteome-wide analysis of protein overabundance. Panel A: TFNseq schematic.} A poly-clonal library of knockout mutants is generated by the transformation of ADP1 with DNA mutagenized by transposon insertions. The library is proliferated on selective media and sequential fractions are collected.  The relative-abundance trajectories of mutants are determined by mapping transposon insertion sites by sequencing. 
 \textbf{Panel B: TFNseq-trajectory analyses for five mutant strains.}  Each mutant  trajectory is well fit by one of the three trajectory models. As expected, the no-effect model is selected for the non-essential gene \textit{recF}. For the other four essential genes, the overabundance model is selected. The dotted line represents the arrest time for each mutant.
 \textbf{Panel C: Overabundance varies by orders of magnitude between essential proteins.} The protein overabundance is inferred from the arrest time using Eq.~\ref{eqn:mutantdilationtime}. Sufficient expression genes have overabundance $o=1$, while overabundant genes vary from $o>1$ to very large overabundance ($o> 100$).  
\textbf{Panel D: Overabundance is large for low-expression essential proteins.}  The measured message-number-overabundance pairs are shown for essential genes (including estimated gene density.) The smoothed experimental data is shown in blue (with experimental uncertainty.)  
      The RLTO model (red) predicts that overabundance grows rapidly as the transcription level is reduced. The RLTO model qualitatively captures the trend of the data (blue); however, it appears to underestimate the measured overabundance for intermediate expression genes. 
      \label{fig:ProteomewideDepletion}}
\end{figure*}

\idea{TFNseq determines overabundances genome-wide.} To test the signature expression-dependent overabudance prediction of the RLTO model, we now transition to a genomic-scale analysis.
%
The Manoil lab  developed a TFNseq-approach to knockout-depletion experiments for targeting all genes simultaneously in \textit{A.~baylyi} \cite{Gallagher:2020sp}. In short: A genomic library was prepared and mutagenized using a transposon carrying the Km$^{\rm R}$ allele. The resulting DNA was then transformed into \textit{A.~baylyi}. The transformants were propagated on selective liquid media 
and fractions collected every two hours from which genomic DNA was extracted. The transposons were then mapped using Tn-seq to generate the relative abundance trajectory for each mutant \cite{Gallagher:2020sp}. (See Fig.~\ref{fig:ProteomewideDepletion}AB.) We then analyzed each mutant trajectory statistically using three competing growth models: no-effect, sufficiency, and overabundance, using two successive null-hypothesis tests. (See Supplementary Material Sec.~\ref{sec:stattraj}.) For each mutant $i$ described by the overabundance model, the TFNseq experiment measures a growth arrest time $T_i$ and the corresponding target protein overabundance:
\begin{equation}
o_i = \exp(k_0 T_i), \label{eqn:mutantdilationtime}
\end{equation}
where $k_0$ is the wild-type growth rate. (See Supplementary Material Sec.~\ref{sec:stattraj}.)


To test the consistency of this TFNseq approach with imaging-based knockout-depletion measurements, we focused first on the analysis of the mutants \textit{dnaA}, \textit{dnaN}, \textit{ftsN}, and \textit{murA}. 
As shown in Fig.~\ref{fig:ProteomewideDepletion}B, the trajectories for \textit{dnaA}, \textit{murA}, \textit{ftsN}, and \textit{dnaN} show an unambiguous step-like change in growth dynamics:   The no-effect trajectory model (null hypothesis) are rejected with p-values that are below machine precision, and the sufficiency trajectory model is also rejected with $p<10^{-4}$ for all genes. 
In Tab.~\ref{tab:overab}, we compare protein overabundances determined by imaging- and sequencing-based approaches. These numbers are qualitatively consistent. For instance, the single-cell analysis of \textit{dnaA} mutant shows a nearly immediate phenotype by imaging (\textit{i.e.}~cell filamentation). (See Supplementary Sec.~\ref{sec:dnaA}.) Likewise, the TFNseq-approach finds an overabundance of 1.0, meaning that protein expression is sufficiency. On the other hand, all three of the other mutants (\textit{murA}, \textit{ftsN}, and \textit{dnaN}) are found to have very large overabundances, and are roughly comparable. Finally, a representative non-essential gene (\textit{e.g.}~\textit{recF}) shows no effect. These results support the use of the TFNseq approach to analyze protein overabundance genome wide.

\idea{Many essential proteins have vast overabundance.}  To determine the protein overabundance genome-wide, we analyzed the knockout-depletion trajectories  for all genes in \textit{A.~baylyi}. (See Fig.~\ref{fig:ProteomewideDepletion}BCD.)
Our analysis showed that the vast majority ($90\%$) of genes annotated as non-essential were classified as having \textit{no effect} and 10\% of non-essential genes had measurable growth defects. (See Supplementary Material Fig.~\ref{fig:pie}.) 
The most severe growth defect in non-essential annotated genes were observed for the genes \textit{gshA} and \textit{rplI}. 
For essential genes, all mutants were observed to have growth defects, as anticipated; however, only 31\% of essential proteins were classified as \textit{sufficient}, corresponding to an immediate change in growth rate. 
Notable genes in this category include ribosomal proteins RpsQ and RpsE, ribonucleotide reductase subunits NrdA and NrdB, and ATP synthase subunits AtpA and AtpD.
However, as predicted by the RLTO model, the majority of essential proteins (69\%), were classified as \textit{overabundant}, meaning that they required significant dilution before a growth rate change was detected.  Fig.~\ref{fig:ProteomewideDepletion}D shows a histogram of essential gene overabundances. 

\idea{Low-expression genes are highly overabundant.}
To understand the overall significance of overabundance in a typical biological process, we determined the  median essential protein overabundance: 7-fold.
To understand the significance of overabundance from the perspective of the metabolic load, we also determine the mean protein overabundance, weighted by the expression level: 1.6-fold.  These two superficially-conflicting statistics emphasize a key predicted regulatory principle: overabundance is high for low-abundance proteins; however, it is close to unity for the high-abundance proteins, which constitute the dominant contribution to the metabolic load.

To explicitly test the predicted relation between protein expression and overabundance, we measured the relative abundance of mRNA messages by RNA-Seq for exponentially growing \textit{A.~baylyi} cells.
(See Supplementary Material Sec.~\ref{sec:smEcvsAb}.)
  We  computed the message number (transcripts per gene per cell cycle) for each essential gene. (See Supplementary Material Sec.~\ref{sec:smmessagenum}.) Fig.~\ref{fig:oaum} compares measured message numbers and overabundances for all essential genes with the prediction of the RLTO model. 
  
  As predicted, the data shows   a clear trend of decreasing overabundance with increasing message number (Fig.~\ref{fig:ProteomewideDepletion}D).  To quantitatively capture this trend, we computed the mean log overabundance over windows of message number (blue curves) to compare the data cloud to the RLTO model predictions. With very few exceptions, high expression genes have extremely low overabundance.  At the other extreme, low expression genes typically have large to very large overabundance as shown by the sharp up-turn of the blue curve as the message number approaches the one-message-rule threshold, a lower threshold on transcription that we recently proposed \cite{Lo2024}. 

\begin{figure}
  \centering
    \includegraphics[width=0.48\textwidth]{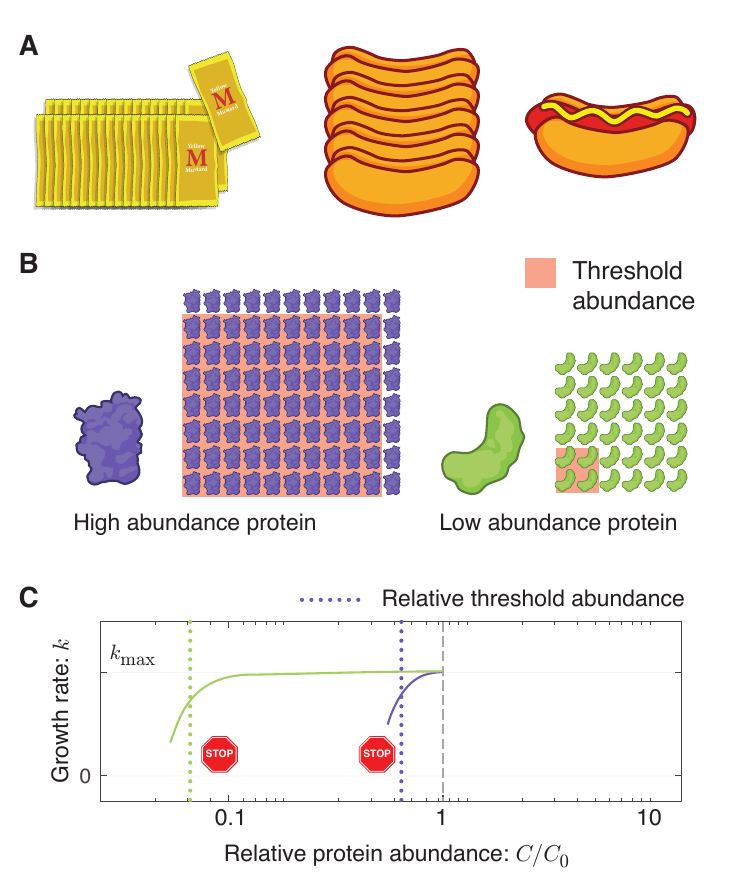}
      \caption{ \textbf{How rate-limited kinetics shapes the fitness landscape. Panel A: An analogy for rate-limited kinetics.} The number of  sausage sandwiches assembled from the pictured ingredients is limited by a single ingredient, the sausages. A depletion of either bun or mustard abundance does not immediately affect the sandwich number.
 \textbf{Panel B: Protein abundance and threshold.} Two essential protein species with different abundances are pictured schematically. The threshold abundance at which each protein becomes limiting is represented by the pink square and the total cellular abundance is represented by the protein array. 
  \textbf{Panel C: Emergent fitness landscape.} A schematic model of the growth rate versus relative protein abundance is shown for the two protein species. The RLTO model predicts that low-abundance proteins (green) have high overabundance, which leads to significant insensitivity to protein depletion. High-abundance protein (purple) are predicted to have small overabundance leads to high sensitivity to protein dilution. The growth rate rapidly decreases with concentration once a species becomes limiting. 
      \label{fig:model}}
\end{figure}

\TocSkip\section*{Discussion}

\idea{The shape of the fitness landscape.} 
Despite some large-scale measurements \cite{Keren:2016pt,Peters:2016jf,Donati:2021kq,Bosch:2021gd}, fundamental questions remain about the structure of the fitness landscape and its rationale \cite{Belliveau:2021pj}. Our measurements reveal that most (69\%) essential proteins show a step-like transition between wild-type and arrested growth below a critical threshold protein abundance. Although asymmetric landscapes have been observed previously (\textit{e.g.}~\cite{Keren:2016pt,Lalanne:2021xs}),
the knockout depletion approach is expected to yield more quantitative results.  For instance, the use of either CRISPRi (\textit{e.g.}~\cite{Peters:2016jf}) or inducible promoters (\textit{e.g.}\ \cite{Lalanne:2021xs}) significantly increases the cell-to-cell variation in protein abundance \cite{Novick:1957oy,Franks:2024uh}, obscuring the features of the fitness landscape. The sharpness of the protein-abundance threshold is manifest in the single-cell analysis  where the progeny begin from a common pool of protein in a single progenitor cell and are therefore not subject to noise (\textit{e.g.}\ Fig.~\ref{fig:singlecell}C).

\idea{The rationale for a threshold abundance.} The observed threshold-like dependence can be rationalized in terms of chemical kinetics: If the protein target is not a rate-limiting reactant in an essential cellular process, then its depletion has no effect on the  rate \cite{NelsonChemicalKinetics,Lo2024}. See Fig.~\ref{fig:model}. We explicitly demonstrate  protein function (\textit{i.e.}~replication) is robust to an order-of-magnitude depletion of replisome protein DnaN; however, for most proteins, we must infer this picture from the growth rate.

\idea{The rationale for overabundance.} Rate-limiting kinetics does not in itself predict vast protein overabundance.  The RLTO model predicts that this feature of the fitness landscape is a consequence of a balance between (i) the metabolic cost of protein expression, which favors minimizing protein abundance, and (ii) robustness to the noise in gene expression \cite{Paulsson:2000xi,Friedman:2006oh}. The model predicts expression-dependent protein overabundance: large  overabundance for low-abundance proteins and small overabundance for high-abundance proteins \cite{Lo2024}.
We show that this signature prediction is observed (Fig.~\ref{fig:ProteomewideDepletion}D). In spite of predicting the genomic-scale trend, there are some significant outliers. We discuss their significance as well as  evidence for the conservation of  overabundance in Supplementary Material Sec.~\ref{sup:disc}

\idea{Biological implications.} 
Many important proposals have been made about the biological implications of noise \cite{Raser:2005we}.  Our work reveals that noise acts to inflate the optimal expression levels of low-expression  proteins and, as a result, significantly increases the metabolic budget for protein, which constitutes 50-60\% of the dry mass of the cell \cite{Lengeler1998}. 
We believe this increased protein budget has cellular-scale implications. For instance, in stress response and  stationary phase, the
 presence of a significant reservoir of overabundant protein  provides critical resources, via protein catabolism, to facilitate the adaptation to changing conditions \cite{Weichart:2003gd,Goldberg:1976rg}.
Protein overabundance may have important implications for individual biological processes as well, including determining which proteins and cellular processes make attractive targets  for small molecule inhibitors (\textit{e.g.}~antibiotics) \cite{Bosch:2021gd}. 
Since overabundance defines the fold-depletion in protein activity required to achieve growth arrest, high-overabundance proteins are predicted to be extremely difficult targets for inhibition.

\idea{Conclusion.}  By combining imaging-, genomic-, and modeling-based approaches, we provide a both a quantitative measurement of the fitness landscape for all essential proteins as well as a clear qualitative and conceptual understanding of the rationale for the observed fitness landscape.  The RLTO model fundamentally reshapes our understanding of the rationale for protein abundance. The model predicts, and experiments confirm, that low-abundance proteins are expressed in vast excess of what is required for growth. Despite the limitations of the experiments, the predicted trend is clearly resolved both at a genomic-scale, using sequencing-based approaches, as well as at the single-cell scale, as observed by microscopy. The rationale for the overabundance strategy is intuitive: Growth requires the robust expression of between five to six hundred distinct proteins. The cell contends with this extraordinary complex regulatory challenge by keeping all but the highest-abundance proteins in vast excess.  


\idea{Data availability.} We include source data files and sequencing data from RNA-Seq experiments to quantify transcription levels. Gene Expression Omnibus (GEO) accession number TBA.  
 
\idea{Acknowledgments.} The authors would like to thank B.~Traxler, A.~Nourmohammad, J.~Mougous,  M.~Cosentino-Lagomarsino, S.~van Teeffelen, C.~Manoil, L.~Gallagher, J.~Bailey, J. Mannik, and S.~Murray.  H.J.C., T.W.L., D.H., and P.A.W.~were supported by NIH grant R01-GM128191 and NSF grant GR046955. K.J.C.~was supported by the Molecular Biophysics Training Program (NIH grant T32GM008268). W.R.W.~was supported by NIH grant R01-AI150041.

\idea{Author contributions:} H.J.C., T.W.L., K.J.C., D.H., and P.A.W.~conceived the research. H.J.C., W.R.W., and P.A.W.~performed the experiments. H.J.C., T.W.L., K.J.C., and P.A.W. performed the analysis. H.J.C., T.W.L., D.H., and P.A.W.~wrote the paper.

\idea{Competing interests:}
The authors declare no competing interests.

\idea{Supplementary Materials}
\\
Supplementary text, Materials and Methods\\
Figs. S1 to S11.\\
Tables S1 to S3.\\
References 35 to 56.\\
Data files S1 to S9. \\
Movies S1 to S12.

\bibliographystyle{science}
\TocSkip\bibliography{Overabundance2}


\onecolumngrid

\newpage 

\beginsupplement

\title{Supplementary material: Protein overabundance is driven by growth robustness}

\maketitle

\onecolumngrid


\tableofcontents


\bigskip
\bigskip

\twocolumngrid

\section{Supplementary Discussion}

\label{sup:disc}

\subsection{Discussion: Are non-essential proteins overabundant?} 

We have focused our analysis on essential genes in the model organism \textit{A.~baylyi} and demonstrated that most essential proteins are overabundant. To what extent is this mechanism generic to non-essential proteins? 
Several arguments support a generic applicability to non-essential genes. 
Our modeling suggests asymmetry rather than explicit growth arrest is the mathematical rationale for the optimality of overabundance \cite{Lo2024}.   We therefore predict that all proteins that increase cell fitness, not just essential proteins, will be overabundant. 
In addition, it is important to emphasize that the annotation of genes as \textit{essential} is contextual. For instance, for \textit{E.~coli} proliferation on lactose, the gene \textit{lacZ} is essential, although non-essential for other carbon sources. As a result, we predict that when expressed, LacZ should be overabundant, consistent with observation \cite{Lambert:2014tp}. 
Finally, The RLTO model also correctly predicts the balance between transcription and translation for all genes, not just essential genes, in eukaryotic cells, suggesting that it should generalize to nonessential genes as well \cite{Lo2024}.

\subsection{Discussion: Limitations of knockout-depletion experiments.}

In spite of the success of the RLTO model in predicting the genomic-scale overabundance trend, there are many significant outliers from this prediction. In considering their significance, it is important to emphasize the flaws both with the knockout-depletion experiments, as well as the RLTO model. With respect to the experiments, the mechanism of growth arrest plays an important role in determining which growth metric most accurately determines the arrest time. Consider the three arrest times measured for the septation-related essential gene \textit{ftsN} in Tab.~\ref{tab:overab}. Due to the absence of strict cell-cycle checkpoints in the bacterial cell, the arrest of the septation process does not immediately arrest cell elongation and replication \cite{Autret:1997zt}. Growth arrest is therefore detected first by the cell-number metric, directly dependent on septation, and later in the other two metrics.

\subsection{Discussion: Limitations of the RLTO model.} 

Likewise, the RLTO model itself has some important limitations. For instance, the model assumes that the dominant contribution to the fitness cost of protein overabundance is metabolic load rather than toxicity \cite{Bolognesi:2018fe}. We have already investigated the consequences of a toxicity-based increase in cost from a model perspective: The qualitative behavior of the model is unchanged; however, the optimal overabundance is reduced by toxicity   \cite{Lo2024}. Motivated by this prediction, we tested whether two classes of proteins, ATPases and enzymes \cite{Bolognesi:2018fe}, that are expected to exhibit toxicity, have lower overabundance. In Supplementary Material Sec.~\ref{sec:tox}, we demonstrate that this predicted trend is observed.
Similarly, the low overabundance of DnaA also provides a second clue about a class of genes that is predicted to have low overabundance: \textit{dnaA} is negatively autoregulated \cite{Menikpurage:2021ax}. Tight regulation can reduce noise, and therefore we hypothesize that tight regulation, and auto-regulation in particular \cite{Nevozhay:2009er}, could therefore reduce the optimal overabundance \cite{Lo2024}. In Supplementary Material Sec.~\ref{sec:reg}, we demonstrate that this predicted trend is also observed in the data. The putative importance both gene-product toxicity and gene regulation in determining the optimal overabundance emphasizes that the RLTO model describes only part of the biology that determines optimal expression levels.

\subsection{Discussion: Is protein overabundance conserved?} 

To what extent is the overabundance of essential genes a conserved mechanism from bacteria, to single-cell eukaryotes, to multicellular organisms? 
As we emphasized above, CRISPRi protein depletions in a wide range of model organisms appear to be consistent with the overabundance hypothesis \cite{McGinness:2006gn,Davis:2011mf,Cameron:2014yt,Peters:2016jf,Liu:2017tm}. Furthermore, we have demonstrated elsewhere that the RLTO model also predicts two other principles of central dogma function (the one-message-rule and load balancing in protein expression) that are observed in eukaryotic cells \cite{Lo2024}. We therefore expect to observe the overabundance strategy in all organisms for low-expression genes \cite{Lo2024}.

\section{\textit{Acinetobacter baylyi} strains, manipulation, and culturing}

Mutant strains were derived from \textit{Acinetobacter baylyi} ADP1 (MAY101)  (the gift of C.\ Manoil) \cite{Barbe:2004ng}. Growth media were LB and M9, a minimal-succinate M9 medium \cite{miller1972experiments}, supplemented with 15 mM sodium succinate, 2 mM magnesium sulfate, 0.1 mM calcium chloride and 1--3 {\textmu}M ferrous sulfate (from sterile 5mM stock, made fresh at least once a month). 
For selective growth, media was supplemented with kanamycin at 20 {\textmu}g/mL.  Cultures were grown at 30$^\circ$C. 

The strains used in the study are summarized in Tab.~\ref{tab:strains}.

\begin{table*}[]
    \centering
\resizebox{1 \textwidth}{!}{    \begin{tabularx}{1\textwidth}{r|r|r|r|r|r|r|X}
\textbf{Short}                 & \textbf{Lab} &                          &                   &                  &                     & \textbf{Selectable}  &  \\
\textbf{name:}                  & \textbf{number:} & \textbf{Organism:}       & \textbf{Genotype:} & \textbf{Source:} & \textbf{Stability:} & \textbf{marker:}    & \textbf{Description:} \\
\hline
\hline
wild-type             & \#1139 & \textit{A.~baylyi}     &  ADP1                                         & Ref.~\cite{Barbe:2004ng} & Stable & --- & Wild-type strain. \\
\rowcolor{blue!10} YdnaN                 & \#1545 & \textit{A.~baylyi}     &  ADP1 \textit{dnaN}::\textit{YPet-dnaN}        & This study.              & Stable & --- &  The beta clamp (DnaN) is replaced by the fluorescent fusion \textit{YPet-dnaN} at the endogenous locus. \\
\hline
$\Delta$\textit{IS}   & N.A.\textsuperscript{\textdagger} & \textit{A.~baylyi}     & \makecell[tr]{ ADP1 \\ \textit{dnaN}::\textit{YPet-dnaN} \\ \textit{ACIA0320-0321}::\textit{kan} }    & This study.              & Stable & Km\textsuperscript{R} &  This is a control strain where non-essential genes, corresponding to an IS element, are knocked out from the YdnaN strain.  Even through the strain is stable, it is re-transformed in each knockout-depletion experiment. Transformed strain has a wild-type growth phenotype.\\
\rowcolor{blue!10}$\Delta$\textit{dnaA}  & N.A.\textsuperscript{\textdagger} & \textit{A.~baylyi}     & ADP1 \textit{dnaA}::\textit{kan}        & This study.              & Unstable & Km\textsuperscript{R} &  DnaA is an essential cell-cycle regulator. This strain must be re-transformed in each knockout-depletion experiment. Transformed strain is wild-type. \\
$\Delta$\textit{dnaN}  & N.A.\textsuperscript{\textdagger} & \textit{A.~baylyi}     & ADP1 \textit{dnaN}::\textit{kan}        & This study.              & Unstable & Km\textsuperscript{R} &  The beta clamp (DnaN) is an essential component of the replisome. This strain must be re-transformed in each knockout-depletion experiment. Transformed strain is wild-type. \\
\rowcolor{blue!10}$\Delta$\textit{YdnaN}  & N.A.\textsuperscript{\textdagger}& \textit{A.~baylyi}     & ADP1 \textit{YdnaN}::\textit{kan}        & This study.              & Unstable & Km\textsuperscript{R} &  The beta clamp (DnaN) is an essential component of the replisome. This strain must be re-transformed in each knockout-depletion experiment. Transformed strain is YdnaN (not wild-type). \\
$\Delta$\textit{murA}  & N.A.\textsuperscript{\textdagger}& \textit{A.~baylyi}     & ADP1 \textit{murA}::\textit{kan}        & This study.              & Unstable & Km\textsuperscript{R} &  The gene product of \textit{murA} is UDP-N-acetylglucosamine 1-carboxyvinyltransferase, an essential protein in synthesizing the precursors of cell wall synthesis. This strain must be re-transformed in each knockout-depletion experiment. Transformed strain is wild-type. \\
\rowcolor{blue!10}$\Delta$\textit{ftsN}  & N.A.\textsuperscript{\textdagger}& \textit{A.~baylyi}     & ADP1 \textit{ftsN}::\textit{kan}        & This study.              & Unstable & Km\textsuperscript{R} &  The gene product of \textit{ftsN} is essential cell division protein FtsN. This strain must be re-transformed in each knockout-depletion experiment. Transformed strain is wild-type. \\
    \end{tabularx}}
    \caption{ \textbf{Summary of strains used in this study.} The \textit{short name} describes the nomenclature of the strains as described in the text. \textdagger Strain re-created by transformations in each knockout-depletion experiment are not stable and therefore are not assigned a \textit{lab strain number} and, due to their instability, cannot be distributed.
    \label{tab:strains}}   
\end{table*}

\subsection{Methods: Construction of deletion mutations}

We generated deletion mutants by transformation of linear DNA fragments, constructed by PCR using extension overlap \cite{Bailey:2019tp}. A homologous overlap of $\sim$2 kb flanking target genes was created that either directly joined (for marker-free deletions) or flanked a kanamycin resistance cassette (for kan-selectable deletions). Unmarked deletions were in-frame.
Kan deletions were constructed from the \textit{kan} gene from plasmid pACYC177 \cite{Chang:1978cz}, in an orientation matching the deleted gene \cite{Bailey:2019tp}. 
PCR reactions were performed using Q5 Polymerase (New England Biolabs) or Phusion HF polymerase (New England Biolabs) and DNA fragments were purified using Qiaquick columns (Qiagen) before transformation.

\subsection{Methods: \textit{A.~baylyi} transformation protocol}

DNA fragments were transformed into \textit{A.~baylyi} cultures prepared as follows.  Cultures were grown overnight in minimal-succinate M9 media with 1 {\textmu}M ferrous sulfate. The  culture was then back diluted 1:5 into fresh medium and grown one hour, shaking at 30$^\circ$C. The DNA fragment was added at 1 {\textmu}g/mL, followed by incubation for 2.5 - 3 hours with shaking, and then plated on selective (for \textit{kan}-deletion cassettes) or non-selective media (for marker-free casettes). Marker-free deletion mutants were identified by screening single colonies by PCR using primers flanking targeted genes. Essential gene kan-marked deletion mutations were selected by plating on protective medium supplemented with 20 {\textmu}g/mL kanamycin. All unmarked and the marked non-essential deletion mutations were verified by PCR.  For essential gene deletions, 0.1--1\% of the cells were transformed, forming microcolonies of cells carrying the deletion.

\subsection{Methods: Construction of YPet-\textit{dnaN} fusion strain}

\label{sec:smypet}

In previous work in \textit{Escherichia~coli} and \textit{Bacillus subtilis}, we visualized fluorescent fusions to the beta sliding clamp (\textit{dnaN}) to study replication \cite{Mangiameli2017,Mangiameli2017b,Mangiameli2018}. The DnaN protein imaging is a convenient tool for studying replication due to its relatively high abundance and the change in its localization, from diffuse (non-replicating cells) to punctate (replicating cells), which serves as a convenient reporter of activity.

To construct a fluorescent fusion to the \textit{A.~baylyi} DnaN protein with a high probability of success, we used the exact same fluorescent protein and linker to that which R.~Reyes-Lamothe had used to construct the \textit{E.~coli} fusion used in our previous work \cite{Reyes-Lamothe:2010nv}. In this approach, we inserted the YPet-linker cassette at the 5' end of the gene. Since the transformation efficiency of \textit{A.~baylyi} is so high, we constructed a marker-free fusion. We screened colonies by both PCR and fluorescence localization.  We then sequenced the mutant \textit{YdnaN} strain to confirm that the desired construct was achieved. (We provide a supplemental file with the sequence.) Like the original \textit{E.~coli} strain, no growth phenotype is observed under experimental conditions.

\section{Growth models for knockout-depletion experiments}

To quantitatively analyze growth in knockout-depletion experiments, we define three nested growth models: (i) \textit{No-Effect}, (ii) \textit{Sufficiency}, and (iii) \textit{Overabundance} models. In our statistical analysis, we will initially treat the No-Effect model as the null hypothesis and the Sufficiency model as the alternative hypothesis. If the null hypothesis is rejected, we will then adopt the Sufficiency model as the null hypothesis and adopt the Overabundance model as the alternative hypothesis.

\subsubsection{No-Effect model}

In the \textit{No-effect model}, the mutant has no effect on the growth rate. The abundance in a log culture will therefore be: 
\begin{equation}
N_{\rm N}(t;N_0) = N_0 e^{k_0t},  \label{eqn:noeff}     
\end{equation}
where $k_0$ is the wild-type growth rate and $N_0$ is the abundance at $t=0$. 

\begin{figure}
  \centering
    \includegraphics[width=0.50\textwidth]{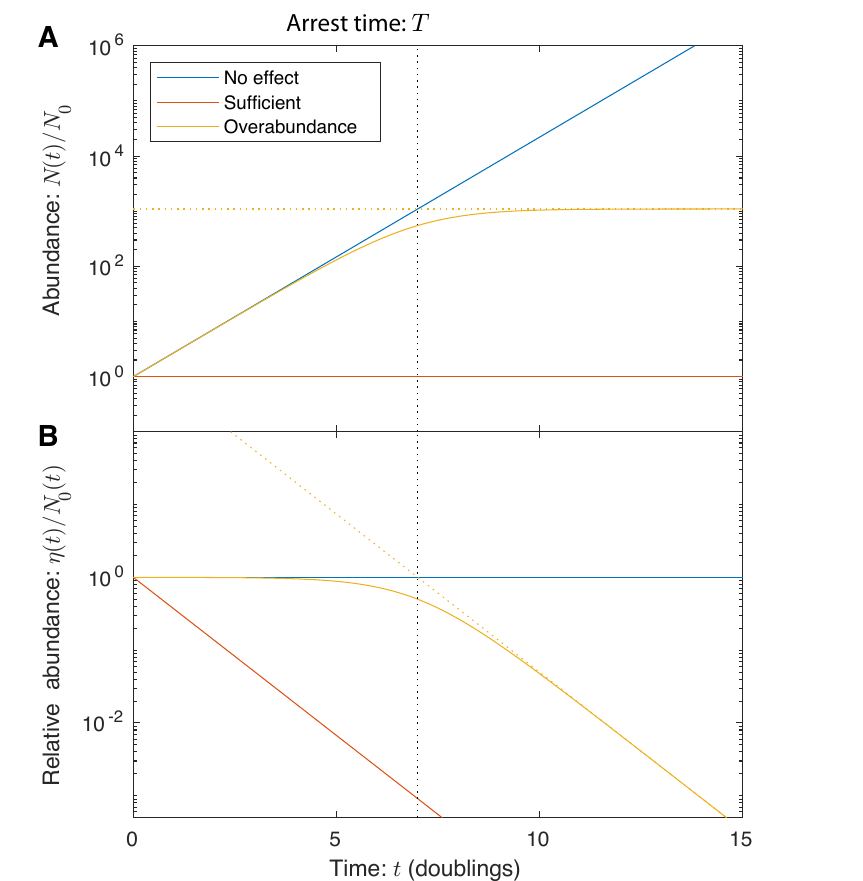}
      \caption{\textbf{Panel A: Mutant abundances for trajectory models.}  Mutants described by the \textit{No-effect model} (blue) grow at the wild-type growth rate. Mutants described by the \textit{Sufficient trajectory model} (red) show an immediate change in growth rate after transformation. Mutants described by the \textit{Overabundant trajectory model} (yellow) grow to the arrest time $T$ (black dotted line) with the wild-type growth rate, before adopting a reduced growth rate of $k=0$.
      \textbf{Panel B: Relative mutant abundances for trajectory models.}   Same as above, but abundances are renormalized by wild-type growth. 
      \label{fig:traj}}
\end{figure}

For modeling the TFNseq trajectories, it is the relative abundance that is measured and we therefore normalize by wild-type growth of the culture, resulting in the relative abundance:
\begin{eqnarray}
\eta_{\rm N}(t;\eta_0) = \eta_0, 
\end{eqnarray}
where $\eta_0$ represents the initial relative abundance. (The relative abundance of the No-effect model is independent of $t$.) Both the abundance $N_{\rm N}$ and relative abundance $\eta_{\rm N}$ are plotted in Fig.~\ref{fig:traj}. Both models depend on a single model parameter and are therefore dimension 1.

\subsubsection{Sufficiency model}

In the \textit{Sufficiency model}, we model the effect of the mutant as immediate. The cell number is assumed to grow at a new unknown rate:
\begin{equation}
N_{\rm S}(t;N_0,k) = N_0 e^{k t},  \label{eqn:suff}
\end{equation}
where $k$ is the new growth rate and $N_0$ is the number of the mutants at $t=0$. For modeling the TFNseq trajectories, it is the relative abundance that is measured, and we therefore normalize by wild-type growth of the culture, resulting in the relative abundance:
\begin{eqnarray}
\eta_{\rm S}(t;\eta_0,\Delta k) = \eta_0 e^{-\Delta k t},  
\end{eqnarray}
where $\Delta k\equiv k_0-k$ is the growth rate reduction of the mutant relative to the wild-type growth rate. Both the abundance $N_{\rm S}$ and relative abundance $\eta_{\rm S}$ are plotted in Fig.~\ref{fig:traj}. Both models depend on two model parameters and are therefore dimension 2. Note that we might na\"ively expect $k = 0$ for essential genes; however, we expect some transient growth due to residual protein levels, and these transients will dominate the fit.

\subsubsection{Overabundance model}

In the \textit{Overabundance model}, we model the effect of the mutant with a delayed arrest time, $T$: the transient growth duration as protein dilutes to the threshold level. For short times, the mutant growth with a wild-type rate:
\begin{equation}
N_{\rm O} =  N_0e^{k_0t},        
\end{equation}
however, at long times we expect growth with a new unknown growth rate $k$:
\begin{equation}
N_{\rm O} = N_0' e^{kt}.
\end{equation}
We initially attempted to use a piecewise function to join these two limits; however, the sparsity of the data and discontinuous slope at the boundary appeared to give rise to fitting artifacts. In addition, the cell-to-cell variation in protein expression smooths out the transition time.  To fix these shortcomings, we adopted an empirical formula with the correct limits, but with a smooth transition at $t = T$:
\begin{equation}
N_{\rm O}(t;N_0,k,T) = N_0e^{k_0 t} \textstyle\frac{e^{\Delta kT}+ 1}{ e^{\Delta kT}+ e^{\Delta k t}},  \label{eqn:over}
\end{equation}
where $\Delta k \equiv k_0-k$ is the loss in growth rate due to the mutation. Modeling the TFNseq trajectories, it is the relative  abundance that is measured, and we therefore normalize by wild-type growth of the culture, resulting in the relative abundance:
\begin{eqnarray}
\eta_{\rm O}(t;\eta_0, k,T) =  \eta_0 \textstyle\frac{e^{\Delta kT}+ 1}{ e^{\Delta kT}+ e^{\Delta k t}},  
\end{eqnarray}
where $\Delta k\equiv k_0-k$ is the growth rate reduction of the mutant relative to the wild-type growth rate. Both the abundance $N_{\rm O}$ and relative abundance $\eta_{\rm O}$ are plotted in Fig.~\ref{fig:traj}. Both models depend on three model parameters and are therefore dimension 3. Note that we might na\"ively expect $k = 0$ for essential genes; however, we expect some transient growth due to residual protein levels, and these transients will dominate the fit.

\section{Imaging-based knockout-depletion experiments}

\label{sec:sm_imaging}

\subsection{Methods: Experimental protocol }

For single-cell imaging-based analyses, cells were imaged proliferating in M9 media supplemented with 2\% low-melt agarose, and in most cases, kanamycin at 20 {\textmu}g/mL.

\subsubsection{Cell preparation for knockout-depletion experiments.} 

\label{sec:cellprep}

The transformation protocol described above was modified as follows: after the 2.5-3 hr incubation with DNA, cells were immediately spotted on selective media pads for imaging. In the knockout-depletion experiments, cells are transformed with knockout cassettes which recombine into the genome, resulting in Km\textsuperscript{R} knockout strains. If transformed cells are transferred to Km\textsuperscript{+} media too quickly, the competent cells do not have sufficient time to integrate the \textit{kan} cassette before growth arrest. If cells are transferred too late, essential proteins are depleted before imaging begins. How do we know transformants after 2.5-3 hr outgrowth are at their initial stages of transient growth? With the 2.5-3 h outgrowth period, many cells still grow slowly (compared to log phase growth) for 10-15 min consistent with the expression of the kanamycin phosphotransferase (the gene product of the \textit{kan} gene) not having reached a sufficiently high level to achieve a resistance phenotype. Furthermore, a significant number of heterogenic progenitors were observed. The presence of these heterogenic progenitor cells is consistent with the 2.5 h outgrowth period representing the typical recombination time for transformants. (See Sec.~\ref{sec:durout} for a discussion of  heterogenic  progenitors.)
\subsubsection{Sample/slide preparation.}  Thin pads were fabricated by melting the agarose (Invitrogen UltraPure\textsuperscript{TM}  LMP Agarose) and casting it between two slides with two layers of lab tape used as a shim to set the height.  After the pad solidified (roughly 10 min), the top slide was carefully removed, and a razor blade was used to trim the pad to form a small square that could be covered with a \#1.5 coverslip. For \textit{E.~coli} imaging, we typically use a pad that matches the size of the coverslip; however, \textit{for \textit{A.~baylyi} imaging, we trim the pad so it is less than 1 cm in width. This added space allows aerobic growth to continue over multiple hours.} Finally, the coverslip is sealed using a hot glue gun.

\subsubsection{Microscopy.} The samples were imaged using a Nikon Eclipse Ti microscope in phase contrast and fluorescence. We imaged through a Nikon 60$\times$ 1.4 NA Phase contrast objective onto a  sCMOS camera (Andor Neo). An environmental chamber maintained the sample at 30$^\circ$C during imaging. For phase imaging, a frame rate of 1 frame / 2 min was used; however, for combined phase and fluorescence imaging we reduced the frame rate to 1 frame / 3 min and 1 frame / 9 min to help reduce bleaching and phototoxicity. (The slowest frame rate was used to resolve the dim YPet-DnaN foci as the protein levels were depleted.) Typically, multiple ($\sim 10$) fields of view were captured simultaneously in each experiment. For fluorescence-based analysis, we mixed in wild-type cells, in addition to fluorescent-fusion cells (1:2), to determine the autofluorescence levels in each experiment. 

\subsubsection{Image processing (cell segmentation) pipeline.} Cell images were processed using the \textit{SuperSegger-Omnipose} package \cite{teresa} by running the \texttt{processExp} command with default settings. Most of the analysis described in the paper was performed from  the \texttt{clist.mat} files generated for each dataset.

\subsection{Methods: Cytometry data analyses }

Imaging-based analysis for protein overabundance was carried out by assessing the transient cell area growth and septation. The three different single-cell analysis approaches are explained below:  \textit{protein abundance}, \textit{area}, and \textit{number analysis}.

\subsubsection{Accessing imaging-based cell cytometry data}
Most of the analysis described in the paper was performed  using the \texttt{clist.mat} files generated for each dataset by the \textit{SuperSegger-Omnipose} package. In particular,  the \texttt{data3D} field provides time-dependent cell descriptors for each cell in each frame, including  \texttt{Rod Length}, \texttt{Area},  \texttt{Fluor1 sum}, and \texttt{Fluor1 background}. These descriptors were the input for our analyses. To characterize cell progeny area of fluorescence, we would generate cell lineage trees and cell progeny IDs using the \texttt{getFamily} command and then sum fluorescence or area over all progeny as a function of time. For instance, this data is shown in Fig.~\ref{fig:WT2}. 
\label{sec:accessclist}

\subsubsection{Protein abundance analysis}

\label{sec:supplfluor}
To test the hypothesis that the targeted protein is depleted while protein-associated function continues for multiple generations, we visualized YPet-DnaN abundance and localization after the protein was knocked out as described in the paper. In short, we constructed a fluorescent fusion at the endogenous locus to make the \text{YdnaN} strain (Sec.~\ref{sec:smypet}), in which the endogenous \textit{dnaN} was replaced by the fusion gene \textit{YPet-dnaN}. In the knockout-depletion experiment, we knocked out the  \textit{YPet-dnaN} gene with the \textit{kan} cassette to form \textit{YPet-dnaN}::\textit{kan}.

To test the protein dilution hypothesis, we measured total progeny fluorescence (the proxy for protein abundance of YPet-DnaN) as a function of time, as the cell progeny proliferated.  The dilution model predicts that the protein abundance should scale with the total progeny area like:
\begin{equation}
C(t) = C(0) \textstyle\frac{A_0}{A(t)},
\end{equation}
where $C(t)$ is the protein concentration at time $t$, $C_0$ is the abundance at time $t=0$, $A_0$ is the progenitor area at time $t=0$, and $A(t)$ is the total area of the progeny at time $t$. In the context of the fusion experiments, the observable is fusion fluorescence, equivalent to an intensity scaling of: 
\begin{equation}
I(t) = I(0) \textstyle\frac{A_0}{A(t)}, \label{eqn:predscaling}
\end{equation}
where $I(t)$ and $I(0)$ are the average pixel intensity of the progeny at time $t$ and the progenitor at $t=0$. Both area $A$ and intensity $I$ are time-dependent quantities available in the \texttt{clist.mat} file. (See Sec.~\ref{sec:accessclist}.) 

Several successive improvements in the experimental design and analysis were required to test the dilution hypothesis. (i) We initially attempted to image cells at the same frame rate as our phase contrast experiment (1 frame/2 min); however, to resolve YPet-DnaN foci after protein depletion, we had to significantly increase the exposure time of the fluorescence images and decrease the frame rate to avoid phototoxicity and bleaching. Although the predicted scaling (Eq.~\ref{eqn:predscaling}) was immediately observable in the data without corrections at short times, more care was required to observe the depletion at long times. (ii) First, we background subtracted to account for the background fluorescence level, computed as the average intensity in each frame outside the cell masks. This correction significantly improved the agreement with Eq.~\ref{eqn:predscaling} at intermediate times, but did not yet account for cellular autofluorescence. (iii) Next, we analyzed a mixture of wild-type and \textit{YdnaN} cells, using the intensity of the wild-type cell in the same microcolony for the background subtraction. This method led to good agreement with Eq.~\ref{eqn:predscaling}  even at long times (Fig.~\ref{fig:dilution}). 

Why was a mixture of wild-type and \textit{YdnaN} cells preferable to imaging the two strains independently? A detailed analysis of single cell intensities  revealed that wild-type cells in close proximity to \textit{YdnaN} cells in the microcolony had higher pixel intensity,  due to the diffuse halo created by the bright \textit{YdnaN} cells. The use of wild-type cells in the same field of view helped correct for the diffuse fluorescent light necessary for the analysis of protein abundance at large depletion times. Cell fluorescence intensities at $t=0$ are used to differentiate between wild-type and \textit{YdnaN} cells.






\subsubsection{Areal growth analysis}

In this section, we develop the statistical model for the analysis of cell-area based growth assays to determine both the model parameters and  the statistical uncertainty of parameters on a per-experiment basis. We provide this development for completeness; \textbf{however, cell-to-cell variation will dominate the reported errors}. 

\iidea{Statistical procedure.} For the imaging-based analyses, we define the following statistical procedure: For the analysis of essential genes, we will fix the asymptotic growth rate $k = 0$. Therefore, the Sufficiency model is now considered the null hypothesis since it is the lowest dimensional model. The first alternative hypothesis is the No-effect model, where the wild-type growth rate $k_0$ is fit in each analysis. If the Sufficiency model is rejected, we then adopt the No-effect model as the new null hypothesis and adopt the Overabundance model as the new alternative hypothesis.

\iidea{Areal growth models.} This growth metric is sensitive to cell elongation (rather than septation). Let $A(t)$ be the observed area of all cells sharing a single progenitor cell. For the areal growth model, we substitute cell area $A(t)$ for the abundance $N(t)$ and  $A_0$ for $N_0$ in Eqs.~\ref{eqn:noeff}, \ref{eqn:suff}, and  \ref{eqn:over}. The models are: 
\begin{eqnarray}
\ln A_{\rm S}(t;A_0) &=& \ln A_0,  \label{eqn:one} \\
\ln A_{\rm N}(t;A_0,k_0) &=& \ln A_0 + k_0t,   \\
\ln A_{\rm O}(t;A_0,k_0,T) &=& \ln A_0 + k_0 t +...\nonumber \\
& & + \ln  \textstyle\frac{e^{k_0T}+ 1}{ e^{k_0T}+ e^{k_0 t}}, \label{eqn:three}  
\end{eqnarray}
where we have substituted $k=0$.

\iidea{Statistical model for areal growth analysis.} We will model the error associated with determining the area of the cells as proportional to cell number or area:
\begin{equation}
\sigma_A \propto A(t).
\end{equation}
This model is consistent with many mechanisms. Rather than fitting a model with a variable error, it is more convenient to introduce a new variable, $a$, with constant error:
\begin{equation}
a(t) \equiv \ln A(t).    
\end{equation}
Since ${\rm d}a = {\rm d}A/A$,  then $\sigma_a = \sigma_A/A$ which leads to an analysis with constant error. 

The Shannon information (minus log likelihood) for the log area in frame $i$ is:
\begin{eqnarray}
h(a_i|{\bm \theta}) &=& \textstyle \frac{1}{2} \ln 2\pi \sigma^2_a + \frac{1}{2\sigma^2_a}[a_i-\mu_a(t_i;{\bm \theta})]^2,   
\end{eqnarray}
where ${\bm \theta}$ represents the parameter vector, $\mu_a$ is the time-dependent mean log area defined by the growth models (Eqs.~\ref{eqn:one}-\ref{eqn:three}). 
For a time series with $i = 1...N$ frames, the total Shannon information is:
\begin{eqnarray}
h(\{a_{i=1...N}\}|{\bm \theta}) &=& \textstyle \frac{N}{2} \ln 2\pi \sigma^2_a + \frac{1}{2\sigma^2_a} S^2,
\end{eqnarray}
which can be formulated as a least-squares minimization where:
\begin{eqnarray}
\Delta a_i &\equiv& a_i-\mu_a(t_i;{\bm \theta}), \label{eqn:delta} \\
S^2({\bm \theta}) &=&  \sum_{i=1}^N \Delta a_i ^2,\label{eqn:S2}
\end{eqnarray}
where $i$ is the frame index.

\iidea{Estimate of error for areal growth analysis.} We will statistically estimate the relative area uncertainty ($\sigma_a$) from the wild-type growth data. The expression for the MLE  for $\sigma_a^2$ is:
\begin{eqnarray}
\hat{\sigma}_{a,\rm MLE}^2 &=& \textstyle \frac{1}{N}S^2(\hat{k}_0,\hat{a}_0), \label{eqn:MLEvar}
\end{eqnarray}
where Eq.~\ref{eqn:S2} is evaluated at the MLE values of the other parameters for the No-effect model. There is one additional improvement to this estimate which is straight forward to implement. It is well known that Eq.~\ref{eqn:MLEvar} is biased from below. We can construct an unbiased estimator by correcting for the complexity of the model for the mean (dimension two) \cite{CoxHink74}: 
\begin{eqnarray}
\hat{\sigma}_{a}^2 &=& \textstyle \textstyle\frac{1}{N-2}S^2(\hat{k}_0,\hat{a}_0), \label{eqn:varUB}
\end{eqnarray}
which we will use for our variance estimator. Note that if only a single mean were fit, the prefactor would be $(N-1)^{-1}$ accounting for the one model dimension; however, since we fit both the slope and the offset, the prefactor is $(N-2)^{-1}$ accounting for the two model dimensions \cite{CoxHink74,BURNHAM2004}. 

From the wild-type growth data, the unbiased estimator for the error for log area (Eq.~\ref{eqn:varUB}) is:
\begin{equation}
\sigma_a = 1.5\times 10^{-3}, \label{eqn:sigmaa}
\end{equation}
or alternatively, this result can be stated in a more intuitive form: There is  a 0.15\% error in the cell area.

\iidea{Application to observed data.} To determine the model parameters (Eq.~\ref{eqn:MLE}), we will minimize the Shannon information (Eq.~\ref{eqn:ShannonInfoNum}) numerically, by a least-squares minimization of  Eq.~\ref{eqn:delta}. We estimate the Fisher information using the resulting Jacobian from the least-squares minimization:
\begin{equation}
\hat{I} \equiv \textstyle\frac{1}{\sigma^2_a}JJ^T,    
\end{equation}
where the Jacobian matrices $J$ are contracted over the frame index and $\sigma_a$ is given by Eq.~\ref{eqn:sigmaa}. The parameter uncertainties are then estimated from the Fisher information (Eq.~\ref{eqn:uncertainty}).
Although Eq.~\ref{eqn:uncertainty} accounts for the statistical uncertainty in the parameters, it does not account for the cell-to-cell variation. We found that this cell-to-cell variation was dominant. We therefore cite this cell-to-cell variation-based uncertainty. For the p-value calculations (Eq.~\ref{eqn:pval}), we compute the test statistic $\lambda$ (Eq.~\ref{eqn:teststat}) from the differences between residual norms for the null and alternative hypotheses:
\begin{equation}
\lambda = \textstyle \frac{1}{2 \sigma_a^2}(S^2_0-S^2_1),    
\end{equation}
where  $\sigma_a$ is given by Eq.~\ref{eqn:sigmaa}, and the residual norms for model I (the null (0) or the alternative (1) hypotheses) are defined in Eq.~\ref{eqn:S2}.

\subsection{ Cell-number growth analysis }
\label{sec:cellnuman}

In this section, we develop the statistical model for the analysis of cell-number based growth assays to determine both the model parameters and  the statistical uncertainty of parameters on a per-experiment basis. We provide this development for completeness; \textbf{however, cell-to-cell variation will dominate the reported errors}. 

\iidea{Statistical procedure.} For the imaging-based analyses, we define the following statistical procedure: For the analysis of essential genes, we will fix the asymptotic growth rate $k = 0$. Therefore, the Sufficiency model is now considered the null hypothesis since it is the lowest dimensional model. The first alternative hypothesis is the No-effect model where the wild-type growth rate $k_0$ is fit in each analysis. If the Sufficiency model is rejected, we then adopt the No-effect model as the new null hypothesis and adopt the Overabundance model as the new alternative hypothesis.

\iidea{Cell-number growth models.} For the cell-number growth model, we use Eqs.~\ref{eqn:noeff}, \ref{eqn:suff}, and  \ref{eqn:over}. The statistical models depend on the growth rates as function of time for model $I$, which we define as:
\begin{equation}
k_I = \textstyle \frac{\partial}{\partial t} \ln N_I(t;\theta_I),     
\end{equation}
where $N_I$ is the cell abundance in model $I$ at time $t$. The growth rates for the respective models are:
\begin{eqnarray}
k_N(t;k_0) &=& k_0,\\
k_S(t) &=& 0,\\
k_O(t;k_0,T) &=& k_0 \cdot [1+e^{k_0 (t-T)}]^{-1}, \label{eqn:tmp}
\end{eqnarray}
where Eq.~\ref{eqn:tmp} interpolates between the initial growth rate $k_0$ and final growth rate $k=0$ at time $T$.

\iidea{Deriving the Shannon information.} Consider an experiment in which images are taken with a high frame rate, where the time duration between frames is $\delta t$. Let the frame number be denoted $I=1...m$ and the number of cells in each frame $N_I$. Let the model for cell growth be formulated such that the growth rate at time $t_I$ is: 
\begin{equation}
k_I = k(t_I;\theta),     
\end{equation}
where $\theta$ represents a parameter vector. In this analysis, we with model cell division as a Markovian process where:
\begin{equation}
\dot{N} = kN,
\end{equation}
which is to say that we will ignore the internal state of cells. For instance, at time $t$, cells have the same rate of division, irrespective of cell age.

In this model, the number of cell divisions $n_I$ that occur over the short time interval $\delta t$ is Poisson distributed:
\begin{equation}
q(n_I|\mu_I) = \textstyle \frac{\mu^{n_I}_I}{n_I!}e^{-\mu_I}, \label{eqn:poiss} 
\end{equation}
where
\begin{equation}
\mu_I \equiv \textstyle \delta t N_I k(t_I;\theta), \label{eqn:mu_I} 
\end{equation}
is the mean number of divisions. 

We now compute the Shannon information associated with the entire experiment:
\begin{equation}
h(\{N_I\}_{I=1...m}|\theta) = -\sum_{I=1}^m \ln q(n_I|\mu_I).
\end{equation}
Substituting Eqs.\ref{eqn:poiss} and \ref{eqn:mu_I}, the equation is simplified to:
\begin{equation}
h = \sum_{I=1}^m \delta t N_I k_I - \sum_{I\in \rm Div} n_I\ln \delta t N_I k_I + \sum_{I\in \rm Div} \ln n_I!, \label{eqn:ShannonInfoNum}
\end{equation}
where Div represents the frames immediately preceding division. For instance, if there is one cell at frame 5 and two cells at frame 6, Div $= \{5\}$.



\iidea{Application to observed data.} To determine the model parameters (Eq.~\ref{eqn:MLE}), we will minimize the Shannon information (Eq.~\ref{eqn:ShannonInfoNum}) numerically, and determine the Hessian at the optimal parameter values to estimate the Fisher information:
\begin{equation}
\hat{I}_{ij} = H_{ij},
\end{equation}
where $H$ is the Hessian matrix. The parameter uncertainties are then estimated from the Fisher information (Eq.~\ref{eqn:uncertainty}).
Although Eq.~\ref{eqn:uncertainty} accounts for the statistical uncertainty in the parameters, it does not account for the cell-to-cell variation. We found that this cell-to-cell variation was dominant. We therefore cite this cell-to-cell variation-based uncertainty. For the p-value calculations (Eq.~\ref{eqn:pval}), we compute the test statistic $\lambda$ (Eq.~\ref{eqn:teststat}) from the differences in the Shannon information (Eq.~\ref{eqn:ShannonInfoNum}).

\subsection{Results: Imaging-based analyses }

\label{sec:smTFNseq}

\subsubsection{Some progenitors have heterogenic progeny}

\label{sec:durout}

Heterogenic progenitors are progenitor cells that are observed to have progeny with two distinct heritable phenotypes: the Km\textsuperscript{R} knockout phenotypes and the Km\textsuperscript{S} wild-type phenotype. For instance, in the $\Delta$\textit{murA} knockout-depletion experiments, progenitors were observed with one daughter whose progeny proliferated for multiple generations on Km\textsuperscript{+} media before lysing, the  knockout phenotype, and whose other daughters proliferated for a short period but maintained wild-type morphology. The maintenance of the  wild-type morphology suggested that the cells were \textit{murA}\textsuperscript{+} Km\textsuperscript{S}. How were these cells able to proliferate while other Km\textsuperscript{S} cells immediately arrested? 

We hypothesize that since both cells had the same progenitor, recombination occurred in the mother cell, after the \textit{murA} gene was replicated, leading to one wild-type chromosome and one $\Delta$\textit{murA} chromosome. The transient growth of the wild-type cells was the result of overabundance of the \textit{kan} gene product APH(3')II being expressed before cell division in the original mother cell.  

Heterogenic progenitor cells appeared frequently for \textit{dnaN} knockout-depletion experiments, presumably because of the location of \textit{dnaN} in the immediate vicinity of the origin, resulting in early replication. In these experiments, an additional test of the heterogenic progenitor hypothesis was possible due to the fluorescent labeling of the target protein. Cells that arrested early with the wild-type morphology showed no protein depletion; whereas cells that displayed the mutant phenotype (filamentation) showed depleted YPet-DnaN levels.

\begin{table*}[]
    \centering
\resizebox{1 \textwidth}{!}{    \begin{tabularx}{1.2\textwidth}{l|rrrr|rrrr|rr}
 \multicolumn{1}{l|}{} & \multicolumn{3}{l}{\textbf{Area} (Cell elongation dependent) } & \multicolumn{1}{r|}{Overabun-}  & \multicolumn{3}{l}{\textbf{Cell-number} (Cell septation dependent) } & \multicolumn{1}{r|}{Overabun-}  & Number  & Number of  \\
                 & Model     &   Growth rate:     & Arrest time: & dance: & Model     &   Growth rate:     & Arrest time: & dance: & of cells: & progenitors:\\
\textbf{Gene:}    & selected: &   $k$ (hr$^{-1})$  & $T$ (hr)  & $\log_{10} o$            & selected: &   $k$ (hr$^{-1})$  & $T$ (hr)  & $\log_{10} o$             & $N_C$ & $N_P$\\
   \hline  
   \hline
   \textit{IS}(Wild-type) & No-effect     &   $0.925\pm0.005$      & NA          & NA       & No-effect     & $1.04\pm0.14$  & NA           &   NA        &   60 & 1    \\
   \textit{dnaA}          & Overabundance &   $1.25\pm0.02$ & $1.2\pm0.1$ & $0.7\pm0.1$ & Sufficiency   & $1.04$         & $0.0\pm0.3$  &   $0.0\pm0.2$  &   4 & 4    \\
   \textit{dnaN}          & Overabundance &   $1.02\pm0.05$ & $4.5\pm7.7$ & $2.0\pm 3.0$ & Overabundance & $0.88\pm0.07$  & $3.8\pm0.1$  &   $1.4\pm0.1$  & 134 & 8    \\
   \textit{ftsN}          & Overabundance &   $0.78\pm0.06$ & $5.2\pm0.3$ & $1.8\pm0.2$ & Overabundance & $1.12\pm0.25$  & $1.3\pm0.4$  &   $0.6\pm0.2$  &  19 & 5    \\
   \textit{murA}          & Overabundance &   $0.70\pm0.08$ & $3.6\pm0.4$ & $1.1\pm 0.1$ & Overabundance & $0.96\pm0.24$  & $2.0\pm0.3$  &   $0.9\pm0.2$  &  16 & 4    \\
   \hline
    \end{tabularx}}
    \caption{\textbf{Detailed results from fitting imaging-based knockout-depletion experiments.} The table summarizes the analysis of cell proliferation by two complementary metrics: area and cell-number analyses. These two metrics depend on distinct cellular processes: Growth in cell area is dependent on cell elongation, whereas the proliferation of cell number is dependent on the septation process. We give two metrics for sample size: the number of progenitors ($N_P$) and the total number of cells analyzed ($N_C$), corresponding to progenitor and progeny. The estimated standard error is provided for parameter fits. 
    \label{tab:detailed}}   
\end{table*}

\subsubsection{Wild-type imaging-based analyses}

\label{sec:wt}
\label{sec:IS}

We analyzed two different strains with wild-type growth phenotypes: wild-type cells (\textit{Acinetobacter baylyi} ADP1) and ACIA0320-0321::\textit{kan}. 


\medskip
\noindent
\textit{IS::\textit{kan}.} To generate a reference wild-type growth phenotype, we choose a non-essential gene with no reported phenotype, genes ACIA0320-0321, corresponding to an IS element. The deletion was performed on the YdnaN strain, which shows no growth phenotype under the experimental conditions. We will abbreviate this strain $\Delta$\textit{IS}. We constructed this deletion and measured its growth relative to wild-type on  Km\textsuperscript{-} media, and no growth phenotype was observed. However, even though this strain can be stably maintained (since ACIA0320-0321 is non-essential), we transformed this cassette using the same protocol in knockout-depletion experiments. As expected, a comparable number of transformants were observed using this construct to those targeting essential genes.

\begin{figure*}
  \centering
    \includegraphics[width=1\textwidth]{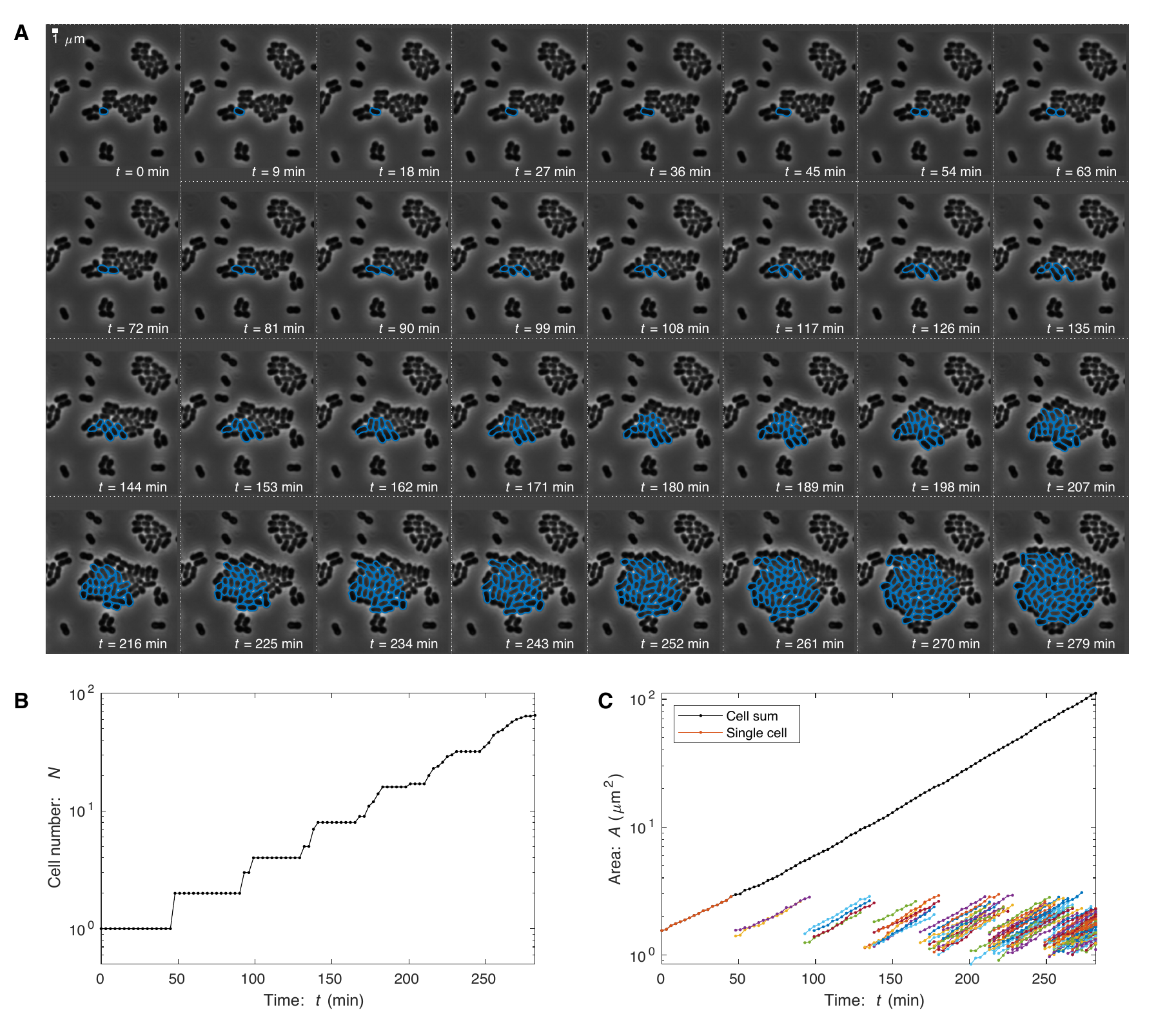}
      \caption{\textbf{Knockout-depletion experiment: IS element (Non-essential).} \textbf{Panel A: Frame mosaic.} In the knockout-depletion experiment, the majority of cells are not transformed and immediately arrest on media supplemented with kanamycin. The lone transformant (\textit{IS}::\textit{kan} (Km$^R$), blue) proliferates normally. Cells were segmented using SuperSegger-Omnipose for quantitative analysis. \textbf{Panel B: Cell number.} The number of transformant progeny as a function of time.  \textbf{Panel C: Progeny area.} Total progeny-cell area as a function of time. Total cell area is plotted with the black-dotted line, while individual cell areas are plotted with color. 
      \label{fig:WT2}}
\end{figure*}

A typical transformant from a knockout-depletion experiment targeting \textit{IS} is shown in Fig.~\ref{fig:WT2} for which six generations of growth are captured. Both the areal (cell-elongation-dependent) and cell-number (septation-dependent) analyses are consistent with the null hypothesis, the \textit{No-effect model}, as expected.   The growth rate was observed to be $k = 0.925 \pm 0.005$ hr$^{-1}$ for the areal analysis and $k = 1.04\pm 0.14$ hr$^{-1}$ for the cell-number analysis.

\iidea{Qualitative phenomenology.} A typical knockout-depletion experiment is shown in Fig.~\ref{fig:WT2}. Panel A shows a frame mosaic. The cells in this dataset show the log-phase growth phenotype of wild-type cells. Both cell number and area show exponential growth. The step-like growth of the cell number reflects the desynchronization of cell division events of the ancestors for a single progenitor.

\iidea{Quantitative analysis.} The null hypothesis (\textit{Sufficiency model}) was rejected in favor of the No-effect model for both the area and cell-number analysis (both p-values under machine precision). The growth rate was observed to be $k = 1.04\pm0.14$ hr$^{-1}$ for the areal analysis and $k = 0.925\pm0.005$ hr$^{-1}$ for the cell-number analysis.


\subsubsection{\textit{dnaA} imaging-based analysis}

\label{sec:dnaA}

\iidea{Annotated gene function.}   DnaA is an essential regulator of the cell cycle and DNA replication initiation in particular.  

\begin{figure*}
  \centering
    \includegraphics[width=1\textwidth]{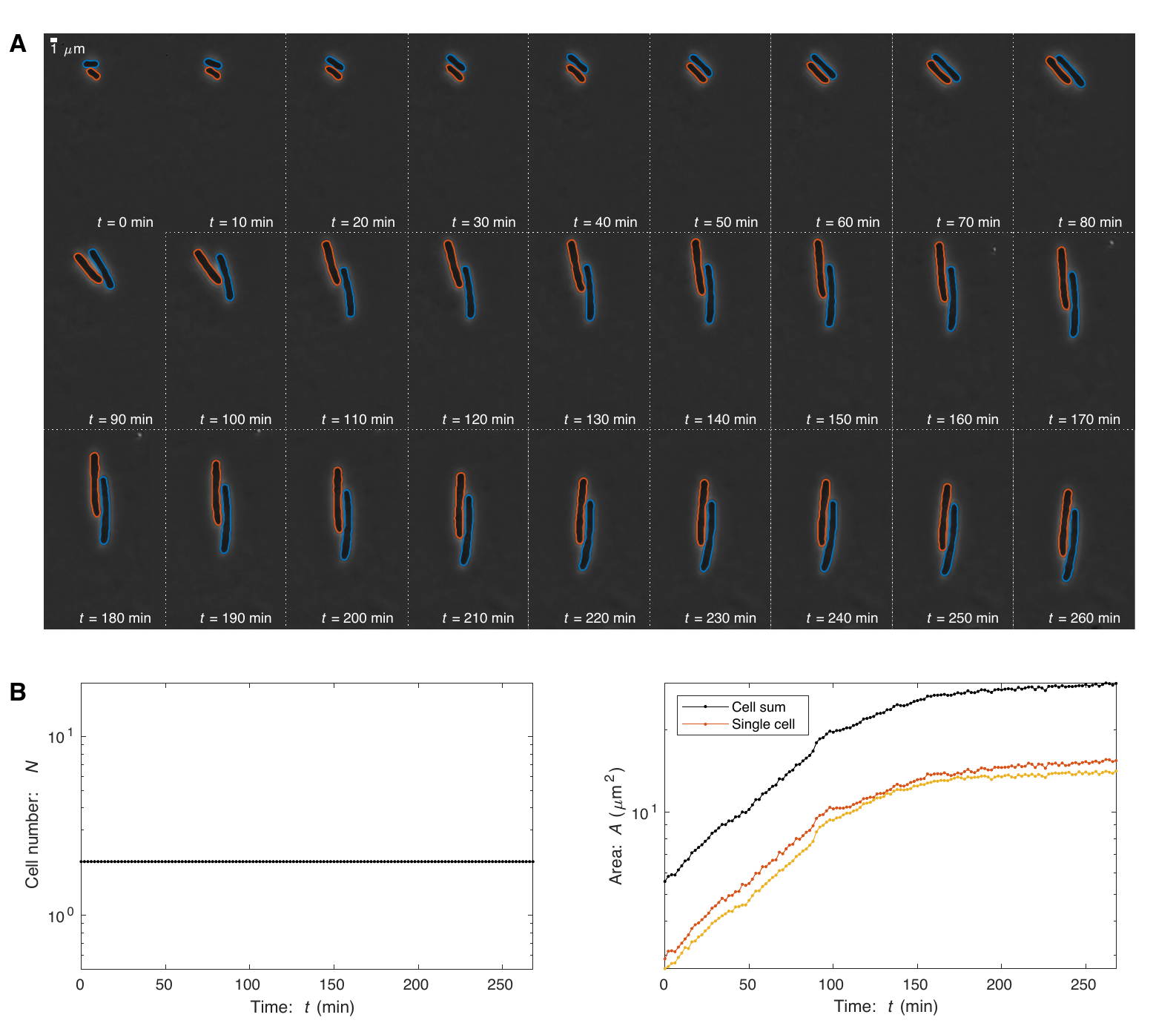}
      \caption{\textbf{Knockout-depletion experiment:} $\Delta$\textit{dnaA}.  \textbf{Panel A: Frame mosaic.} Two transformants (\textit{dnaA}::\textit{kan}(Km$^R$), blue, orange) proliferate. DnaA is an essential regulator of replication initiation.  Its depletion leads to a failure of the chromosome to replicate, and therefore results in cell filamentation.
      Cells were segmented using SuperSegger-Omnipose for quantitative analysis.
      \textbf{Panel B: Cell number.} The number of transformant progeny as a function of time. After transformation, cells fail to divide, consistent with DnaA expression being sufficient rather than overabundant.   \textbf{Panel C: Progeny area.} Total progeny-cell area as a function of time. In spite of the arrest of septation/division, cell areal elongation persists for roughly 120 minutes. 
      \label{fig:dnaA}}
\end{figure*}

\iidea{Qualitative phenomenology.} A typical knockout-depletion experiment is shown in Fig.~\ref{fig:dnaA}. Panel A shows a frame mosaic. The cells in this dataset show the onset of the phenotype, cell filamentation, without undergoing significant growth-induced protein dilution. As a result, the cell number, shown in Panel B, is constant since no divisions are observed.  However, as shown in Panel C, cell elongation continues for roughly 100 min before it begins to arrest. We interpret the metric that shows the earliest arrest to define the overabundance. In this case, since septation is not observed again after transformation, DnaA abundance is consistent with the Sufficiency model.

\iidea{Quantitative analysis.} The null hypothesis (\textit{Sufficiency model}) was rejected in favor of the \textit{Overabundance model} (p-value under machine precision) for the areal analysis. The initial growth rate was observed to be $k = 1.25 \pm 0.02$ hr$^{-1}$ with an arrest time of $T = 1.24 \pm 0.10$ hr. In case of the cell number analysis, we fail to reject the null hypothesis (\textit{Sufficiency model}), indicating that there is no statistical significance to support the alternative hypothesis \textit{No-effect model} (p = 1.0). We used the $\Delta$\textit{IS} wild-type growth rate ($k = 0.925\pm0.005$ hr$^{-1}$) to fit the arrest time: $T = 0.0\pm 0.3$ hr.

\subsubsection{\textit{dnaN} imaging-based analysis}
\label{sec:dnaN}

\iidea{Annotated gene function.}  The gene product of \textit{dnaN} is the $\beta$ sliding clamp (DnaN), which is an essential component of the replisome complex.

\begin{figure*}
  \centering
    \includegraphics[width=1\textwidth]{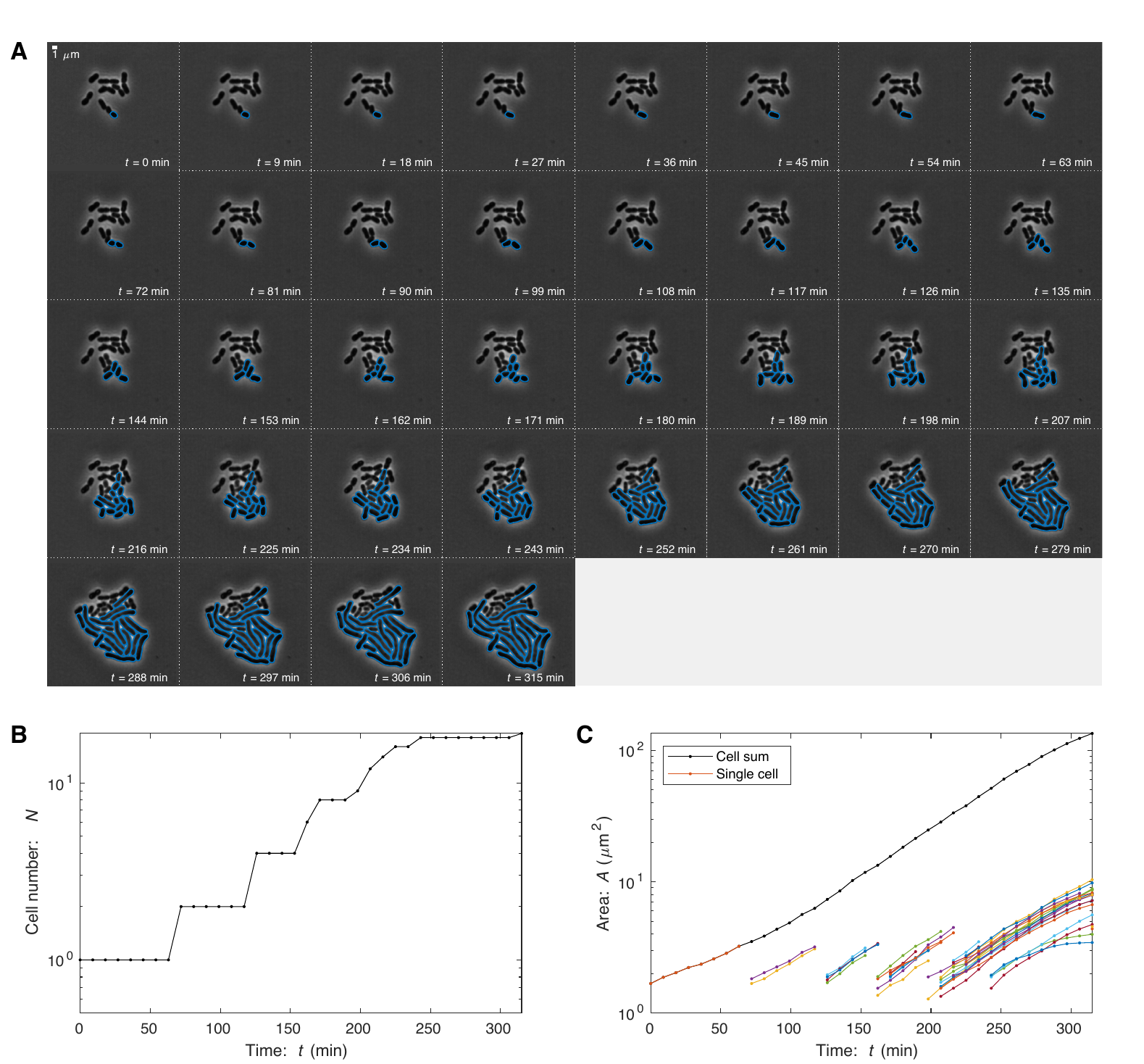}
      \caption{\textbf{Knockout-depletion experiment:} $\Delta$\textit{dnaN}.  \textbf{Panel A: Frame mosaic.} One transformant (\textit{dnaA}::\textit{kan}(Km$^R$), blue) proliferates.  
            The frame mosaic shows a typical imaging-based knockout-depletion experiment. DnaN is the sliding beta clamp, an essential DNA replication protein and a core component of the replisome.   Its depletion leads to a failure of the chromosome to replicate and therefore results in cell filamentation. Cells were segmented using SuperSegger-Omnipose for quantitative analysis. \textbf{Panel B: Cell number.} The number of transformant progeny as a function of time. After transformation, normal growth persists for roughly 240 min, consistent with DnaN expression being overabundant.   \textbf{Panel C: Progeny area.} Total progeny-cell area as a function of time. The areal elongation dynamics persists even are cell division arrests. 
      \label{fig:dnaN}}
\end{figure*}

\iidea{Qualitative phenomenology.} A typical knockout-depletion experiment is shown in Fig.~\ref{fig:dnaN}. Panel A shows a frame mosaic. The cells in this dataset show the onset of the phenotype, cell filamentation, at about 220 min, after multiple rounds of cell division. As a result, the cell number, shown in Panel B, plateaus shortly after the filamentation is observed since the filamentation is a consequence of the failure of the cells to efficiently septate.
However, as shown in Panel C, cell elongation continues, although slowing slightly, throughout the experiment. In this case, since arrest is observed first with respect to septation, we use the arrest of this process to define overabundance.

\iidea{Quantitative analysis.} The null hypothesis (\textit{Sufficiency model}) was rejected for both the area ($p = 8.9 \times 10^{-140}$) and cell-number analysis ($p = 6.0 \times 10^{-19}$). The initial growth rate was observed to be $k = 1.02 \pm 0.05$ hr$^{-1}$ with an arrest time of $T = 4.5 \pm 7.7$ hr for the areal analysis. For cell-number analysis, the initial growth rate was observed to be $k = 0.88 \pm 0.07$ hr$^{-1}$ with an arrest time of $T = 3.8 \pm 0.1$ hr.

\subsubsection{\textit{ftsN} imaging-based analysis}

\label{sec:ftsN}

\iidea{Annotated gene function.}  The gene product of \textit{ftsN} is essential cell division protein FtsN.

\begin{figure*}
  \centering
    \includegraphics[width=1\textwidth]{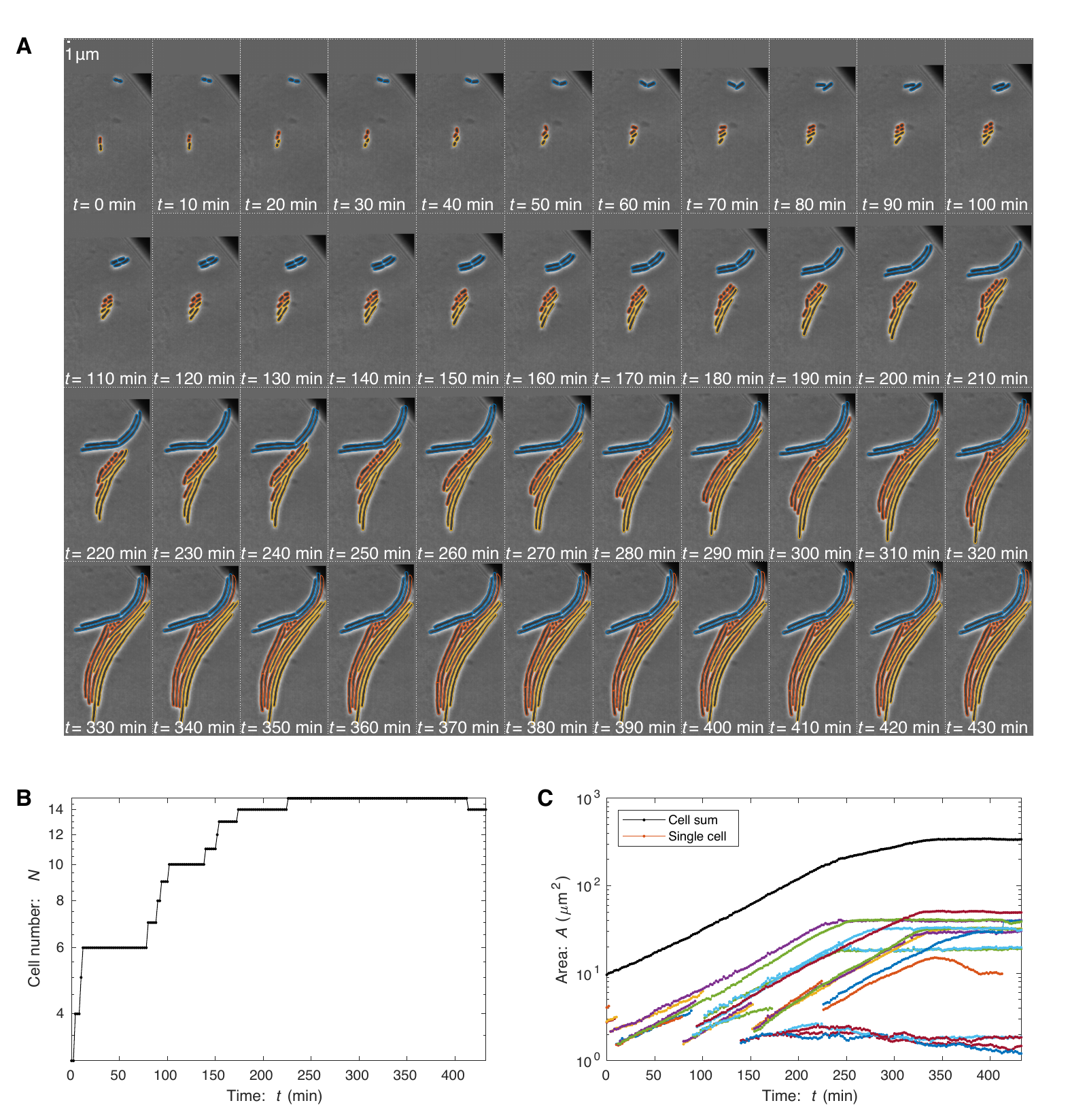}
      \caption{\textbf{Knockout-depletion experiment:} $\Delta$\textit{ftsN}.  \textbf{Panel A: Frame mosaic.} 
      Three transformants (\textit{ftsN}::\textit{kan}(Km$^R$), blue, yellow, orange) proliferate.     FtsN is an essential cell division protein. Its depletion leads to a failure of the cells to septate. Cells were segmented using SuperSegger-Omnipose for quantitative analysis.   \textbf{Panel B: Cell number.} 
      The number of transformant progeny as a function of time. After transformation, normal growth persists for roughly 150 min, consistent with FtsN expression being overabundant.   \textbf{Panel C: Progeny area.} Total progeny-cell area as a function of time. The areal elongation persists even after cell division arrests.
      \label{fig:ftsN}}
\end{figure*}

\iidea{Qualitative phenomenology.} A typical knockout-depletion experiment is shown in Fig.~\ref{fig:ftsN}. Panel A shows a frame mosaic. The cells in this dataset show the onset of the phenotype: the failure to septate, at roughly 150 minutes, after several rounds of division. As a result, the cell number, shown in Panel B, plateaus shortly after 150 min as a consequence of the failure of the cells to efficiently septate.
However, as shown in Panel C, cell elongation continues, although slowing slightly, to roughly 220 min. 

\iidea{Quantitative analysis.}  The null hypothesis (\textit{Sufficiency model}) was rejected for both the area (p-value under machine precision) and cell-number analysis ($p = 1.7 \times 10^{-7}$). The initial growth rate was observed to be $k = 0.78 \pm 0.06$ hr$^{-1}$ with an arrest time of $T = 5.2 \pm 0.3$ hr for the areal analysis. For cell-number analysis, the initial growth rate was observed to be $k = 1.12 \pm 0.25$ hr$^{-1}$ with an arrest time of $T = 1.3 \pm 0.4$ hr.

\subsubsection{\textit{murA} imaging-based analysis}
\label{sec:murA}

\iidea{Annotated gene function.}  The gene product of \textit{murA} is UDP-N-acetylglucosamine 1-carboxyvinyltransferase, an essential protein in synthesizing the precursors of cell wall synthesis.

\begin{figure*}
  \centering
    \includegraphics[width=1\textwidth]{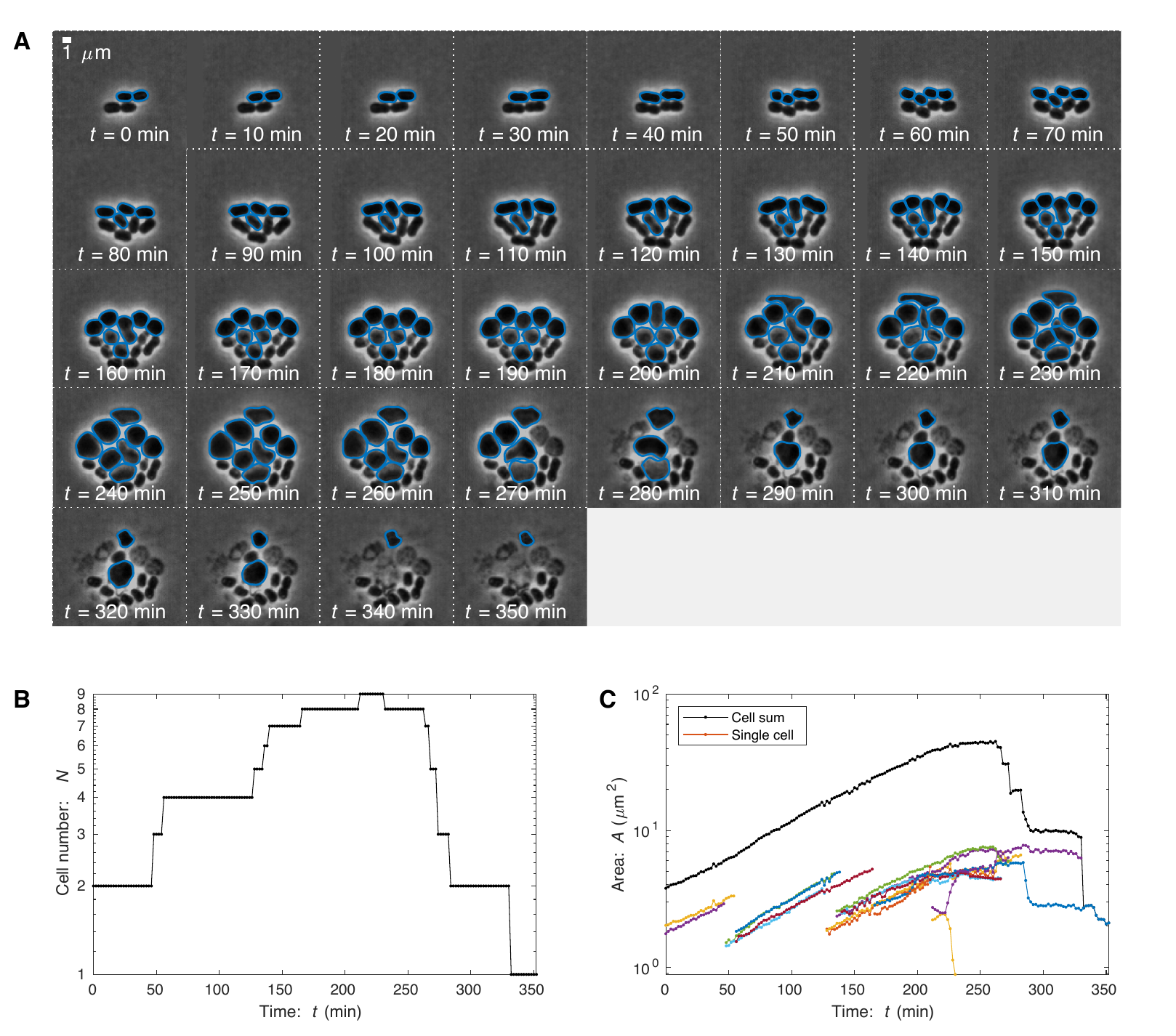}
      \caption{\textbf{Knockout-depletion experiment:} $\Delta$\textit{murA}.  \textbf{Panel A: Frame mosaic.}  Two transformants (\textit{murA}::\textit{kan}(Km$^R$), blue) proliferate.    MurA is an essential enzyme responsible for cell wall precursor synthesis. Its depletion leads to the loss of cell wall integrity, and therefore first the loss of wild-type cell morphology and then cell lysis.   Cells were segmented using SuperSegger-Omnipose for quantitative analysis.   \textbf{Panel B: Cell number.} The number of transformant progeny as a function of time. After transformation, normal growth persists for roughly 200 min, consistent with MurA expression being overabundant.   \textbf{Panel C: Progeny area.} Total progeny-cell area as a function of time. The areal elongation dynamics are largely consistent with the cell number dynamics: Normal growth persists for roughly 200 min.  
      \label{fig:murA}}
\end{figure*}

\iidea{Qualitative phenomenology.} A typical knockout-depletion experiment is shown in Fig.~\ref{fig:murA}. Panel A shows a frame mosaic. The cells in this dataset show the onset of the phenotype: the loss of cell wall integrity, and therefore first the loss of wild-type cell morphology and then cell lysis. Cells begin to lose their wild-type morphology at roughly 120 min, after multiple rounds of cell division. As a result, the cell number, shown in Panel B, plateaus shortly after 150 min as a consequence of the failure of the cells to efficiently septate.
However, as shown in Panel C, cell elongation continues, although slowing slightly, to roughly 200 min. 

\iidea{Supplemental approach.} For this analysis, we did not want to explicitly model cell lysis. Therefore, in our fitting of the cell-number and areal growth curves, we locked the individual cell area at the last value taken immediately preceding lysis. Similarly, we treated cells that had lysed as arrested, not absent. (This fitting-refined data is \textit{not} shown in Fig.~\ref{fig:murA}. The resulting refined data for Panels B and C plateau rather than decease after growth arrest.) 

\iidea{Quantitative analysis.} The null hypothesis (\textit{Sufficiency model}) was rejected for both the area (p-value under machine precision) and cell-number analysis ($p = 1.2 \times 10^{-6}$). The initial growth rate was observed to be $k = 0.70 \pm 0.08$ hr$^{-1}$ with an arrest time of $T = 3.6 \pm 0.4$ hr for the areal analysis. For cell-number analysis, the initial growth rate was observed to be $k = 0.97 \pm 0.24$ hr$^{-1}$ with an arrest time of $T = 2.0 \pm 0.3$ hr.

\section{Statistical analysis of TFNseq trajectories}

\label{sec:stattraj}

\subsection{Methods: Time correction}
Since the mutants transition from lag phase to log phase after transformation, we used a log phase equivalent time for the TFNseq-approach analysis. The corrected sampling times ($t_s$) are estimated from the number of doublings ($D_s$) for the non-essential mutants obtained from TFNseq experiment(\cite{Gallagher:2020sp}): 
\begin{equation}
t_s = D_s*[doubling \;time].      
\end{equation}
For our experiment, the doubling time for ADP1 in M9 at 30$^\circ$C is 37 min.

\subsection{Methods: Defining the likelihood}

We assume that deep-sequencing is well modeled by a Poisson process for which the probability mass function is:
\begin{equation}
p(n|\mu) = \textstyle \frac{\mu^n}{n!}e^{-\mu},
\end{equation}
where $n$ is the number of reads and $\mu$ is the mean-number parameter. For large $n$, we use the normal-distribution approximation:
\begin{equation}
p(n|\mu) \approx \textstyle \frac{1}{\sqrt{2\pi \mu}} \exp \left[ -\textstyle \frac{(n-\mu)^2}{2\mu} \right].
\end{equation}
The total likelihood for sequential observations $n_{1...m}$ at time $t_{1...m}$ is therefore:
\begin{equation}
q(n_{1...m}|\theta_I) = \prod_{i=1}^m  p(n_i|\mu)|_{\mu = N_I(t_i|\theta_I)}, 
\end{equation}
where $N_I$ is one of the trajectory models and $\theta_I$ is the parameter vector for model $I$. The Shannon information is:
\begin{eqnarray}
h(n_{1...m}|\theta_I) &\equiv& -\ln q(n_{1...m}|\theta_I), \\
                      &=& -\sum_{i=1}^m  \ln p(n_i|\mu)|_{\mu = N_I(t_i|\theta_I)}.
\end{eqnarray}



\begin{figure}
  \centering
    \includegraphics[width=0.50\textwidth]{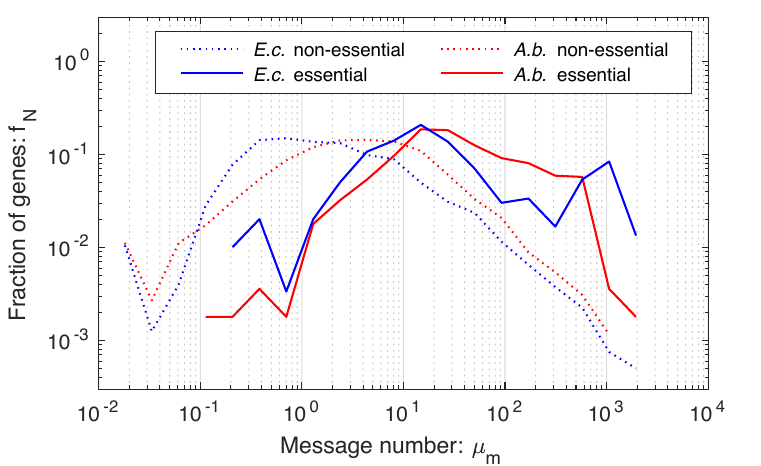}
      \caption{\textbf{Message number distributions for essential and non-essential genes in \textit{E.~coli} and \textit{A.~baylyi}.} Nearly all \textit{A.~baylyi} essential genes are expressed above the one-message-per-cell-cycle threshold.  This distribution of both non-essential and essential genes in \textit{A.~baylyi} is qualitatively similar to that in \textit{E.~coli}, as predicted \cite{Lo2024}.
      \label{fig:onemessage}}
\end{figure}

\begin{figure*}
  \centering
    \includegraphics[width=1.0\textwidth]{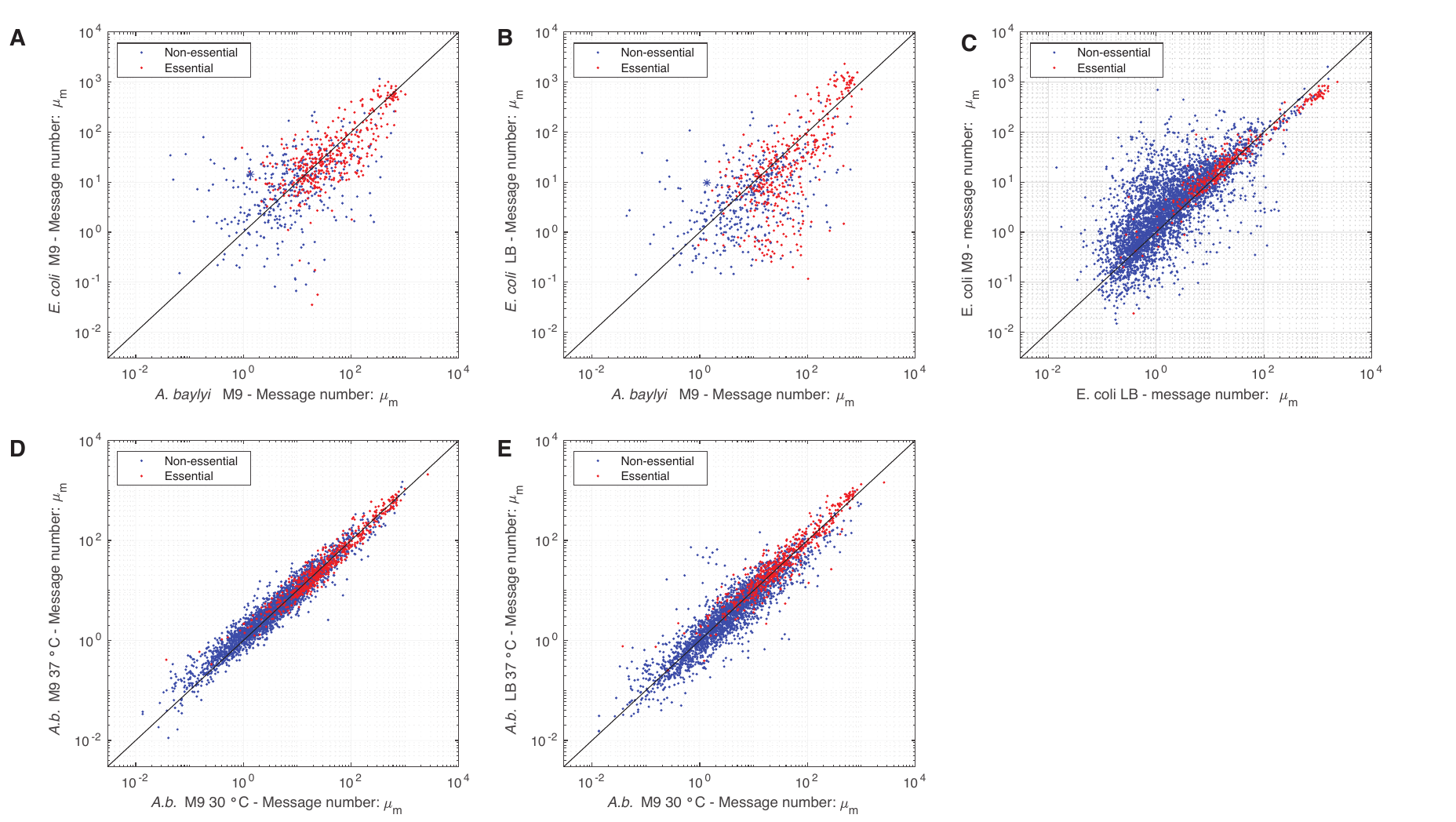}
      \caption{\textbf{Transcriptome comparisons.} \textbf{Panel A: \textit{E.~coli} on M9 versus \textit{A.~baylyi} on M9.}  \textbf{Panel B: \textit{E.~coli} on LB versus \textit{A.~baylyi} on M9.} \textbf{Panel C: \textit{E.~coli} on M9 versus on LB.} \textbf{Panel D: \textit{A.~baylyi} on M9 at 37$^\circ$C versus at 30$^\circ$C.} \textbf{Panel E: \textit{A.~baylyi} on LB at 37$^\circ$C  versus on  M9 at 30$^\circ$C .} Throughout, there is broad consistency between the expression levels (message number) of genes, both between organisms and between conditions. These observations suggest a consistent overall transcriptional program governs gene expression both between organisms and growth conditions.      
      \label{fig:EcolivsAbaylyi}}
\end{figure*}

\begin{figure}
  \centering
    \includegraphics[width=0.47\textwidth]{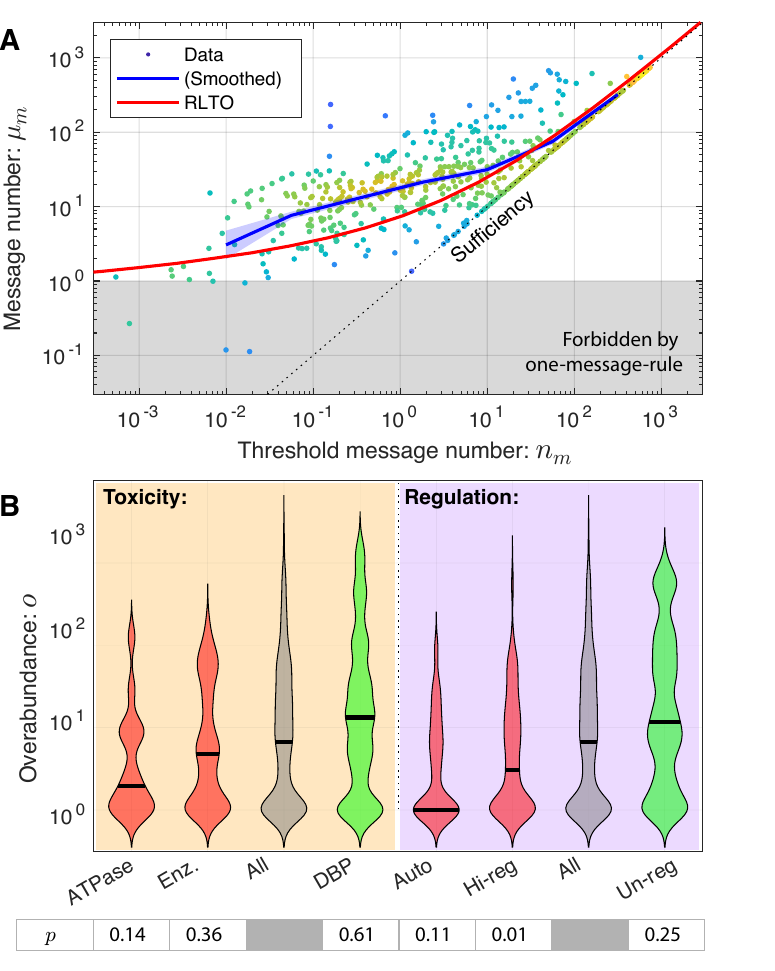}
      \caption{
      \textbf{Panel A: Threshold and message number.} Alternatively, the message number $\mu_m$ is shown as a function of the inferred threshold message number: $n_m\equiv \mu_m/o$. Although it is $\mu_m$ and $o$ that are most directly measurable, the quantity $n_m$ is a more intuitive quantity from a modeling perspective since $\mu_m$ is optimized to maximize fitness at fixed $n_m$ in the RLTO model. \textbf{Panel B: Toxicity and regulation are determinants of overabundance.} We compared the overabundance measurements for six essential gene subgroups to determine whether toxicity and regulation could affect overabundance. Red groups were predicted to decrease overabundance while green groups were expected to increase it. The p-values for the consistency of each distribution with the all gene group is shown below each category. As hypothesized, the data is consistent with both toxicity and regulation decreasing overabundance.
      \label{fig:oaum}}
\end{figure}

\subsection{Methods: Analysis of overabundance for different Gene Ontologies (GO) }
To classify genes,  the gene ontology classifications and terms summarized in Tab.~\ref{tab:go} were used.

\begin{figure}
  \centering
    \includegraphics[width=0.48\textwidth]{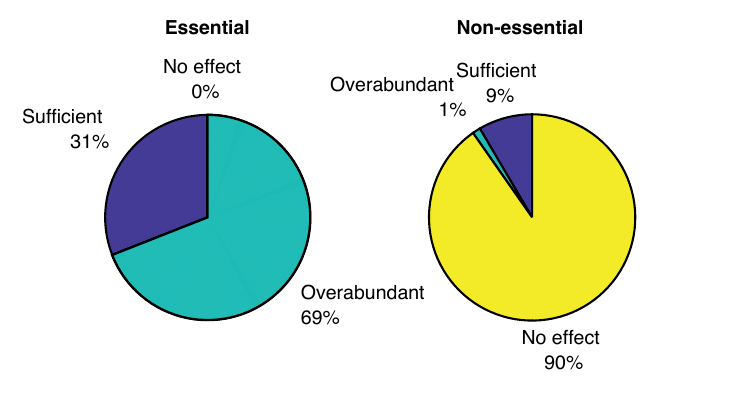}
      \caption{ 
      \textbf{Proteome-wide analysis of proliferation dynamics.} For genes classified as essential, 31\% were best fit by the sufficiency model, while 69\% were best fit by the overabundance  model.  For genes classified as non-essential, 90\% were best fit by the no-effect model, while 10\% showed a detectable reduction in growth rate. 
      \label{fig:pie}}
\end{figure}

\subsection{Results: Toxicity reduces overabundance.} 

\label{sec:tox}
A second key assumption in the RLTO model is that the metabolic cost of transcription and translation are the dominant fitness costs of protein overabundance (\textit{i.e.}~there is no toxicity) \cite{Lo2024}. 
To explore the potential role of toxicity, we generated groups of essential ATPases and enzymes, hypothesizing that these proteins would have higher cost due to excessive activity when overabundant, and a group of DNA-Binding Proteins (DBP), which we hypothesized would have low cost when overabundant. We find that the median overabundance for ATPase genes is 2-fold, and for enzymes more generally 5-fold, compared to 7-fold for all essential genes and 13-fold for DBP. These results are consistent with the hypothesis that toxicity, and in particular ATPase activity, is also a key determinant of overabundance. (See Fig.~\ref{fig:oaum}B.)

\subsection{Results: Regulation reduces overabundance.} 

\label{sec:reg}

Instead, we adopted a hypothesis-driven approach and attempted to construct subgroups of essential genes that violate the underlying assumptions used to formulate the RLTO model.  A key assumption  is that gene expression noise is a consequence of the message number only and is otherwise independent of regulation \cite{Lo2024}. Precise control of expression could lead to a reduction in the optimal overabundance.  
To explore the regulatory hypothesis,  we generated three lists of essential genes: autoregulatory, highly regulated (top 10\% of genes ranked by number of regulators), and un-regulated. If regulation can obviate the need for overabundance, we would expect lower median overabundances in both regulated groups and potentially higher overabundances for the un-regulated group. Consistent with this hypothesis, we find that the median overabundance for autoregulatory genes is 1-fold and for highly-regulated, 3-fold, compared with 7-fold for all essential genes, and 12-fold for un-regulated genes, strongly supporting the hypothesis that tight regulation could reduce the need for overabundance. (See Fig.~\ref{fig:oaum}B.)


\begin{table}
    \centering
\begin{tabularx}{0.49\textwidth}{l|l}
\textbf{Class} & \textbf{Terms} \\
\hline
GO:0006260 & DNA replication \\
GO:0051301 & Cell division \\
GO:0008610 & Lipid biosynthetic process \\
GO:0009252 & Peptidoglycan biosynthetic process \\
GO:0008643 & Carbohydrate transport \\
GO:0006355 & Regulation of DNA-templated transcription \\
GO:0003824 & Catalytic activity \\
GO:0003677 & DNA binding \\
\end{tabularx}   
    \caption{\textbf{Gene ontology classifications and terms.} A summary of  the gene ontology classifications and terms used in the study.
    \label{tab:go}}
\end{table}

\medskip

We analyzed only genes in \textit{A.~baylyi} that had homologues in \textit{E.~coli}. The \textit{E.~coli} classifications were downloaded from EcoCyc database \cite{ecocyc2023}. 

\subsection{Analysis of overabundance for different gene regulatory controls}

 To investigate the effect of transcriptional regulation in determining protein overabundance, we assumed that the regulatory network in \textit{A.~baylyi} is roughly equivalent to that in \textit{E.~coli} which has been much more extensively studied. We used the EcoCyc database \cite{ecocyc2023} to generate a list for each gene $i$ of the list of direct regulators. For each gene, we counted the direct regulators of each gene, then ranked the genes in term of regulator number, and finally we defined the top 10\% of the genes as \textit{highly regulated}. We also generated a list of genes directly inhibited by each gene.
 If a gene directly inhibited itself, we defied the gene as \textit{autoregulatory}.

\section{RNA-Seq analysis of transcription}

\subsection{Methods: RNA-Seq protocol}

\subsubsection{RNA extraction} 

ADP1 RNA was harvested through methods developed by Culviner et al. \cite{Culviner2020} Total RNA was harvested by mixing 1ml of \AB (~0.5 OD) with 110ul of ice-cold stop solution (95\% ethanol and 5\% acid-buffered phenol) and spinning in a tabletop centrifuge for 30 s at 13,000 rpm. The supernatant was flash-frozen and stored at -80{\textdegree}C until RNA extraction is ready. To start RNA extraction, 1ml of heated 65{\textdegree}C was added to the sample. The mixture was shaken at 65{\textdegree}C for 10 min and flash-frozen at -80{\textdegree}C for at least 10 min. The pellets were thawed at room temperature and spun at top speed in a benchtop centrifuge at 4{\textdegree}C for 5 min. The supernatant was collected and added to 400 {\textmu}l of 100\% ethanol. The mixture was passed through  DirectZol spin column (Zymo). The column was washed twice with RNA prewash buffer and once with RNA wash buffer (Zymo). RNA was eluted from the column with 90 {\textmu}l diethyl pyrocarbonate (DEPC)-H2O. Genomic DNA was removed with 4 {\textmu}l of Turbo DNase I (Invitrogen) and supplemented with 10{\textmu}l of 10x Turbo DNase I buffer to a final volume of 100{\textmu}l. The solution was heated to 37{\textdegree}C for 40 min. Then RNA was diluted with 100 {\textmu}l DEPC-H2O, extracted with 200 {\textmu}l buffered acid phenol-chloroform, followed by ethanol precipitation at -80{\textdegree}C for 4 h with 20 {\textmu}l of 3 M sodium acetate (NaOAc), 2 {\textmu}l GlycoBlue (Invitrogen), and 600 {\textmu}l ice-cold ethanol. To pellet RNA, the samples were centrifuged at 4{\textdegree}C for 30 min at 21,000 × g. The pellets were washed twice with 500 {\textmu}l of ice-cold 70\% ethanol, followed by centrifugation at 4{\textdegree}C for 5 min. RNA pellets were then air dried and resuspended in 50 {\textmu}l DEPC-H2O. The yield and integrity of RNA was verified with NanoDrop spectrophotometer, and by running 50 ng of total RNA on a Novex 6\% Tris-buffered EDTA (TBE)-urea polyacrylamide gel (Invitrogen).

\subsubsection{rRNA Depletion} 

rRNA was depleted through the DIY method developed by Culviner et al. \cite{Culviner2020} as well. We used their 21 biotinylated oligonucleotides for \textit{E.~coli}. The selected biotinylated oligonucleotides were synthesized by IDT and resuspended to 100 {\textmu}M in TE buffer (Qiagen). An oligonucleotide mixture was made by mixing equal volumes of each 16S and 23S primers and double volumes of 5S primers. The pooled mixture was diluted with DEPC treated H2O based on the total RNA, using their Excel-based calculator. Using the Excel-based calculator, the calculated volume of Dynabeads MyOne streptavidin C1 beads (ThermoFisher) were washed three times in equal volume of 1x B{\&}W buffer, resuspended in 30 {\textmu}l of 2× B{\&}W buffer and supplemented with 1{\textmu}l of SUPERase-In RNase inhibitor (ThermoFisher). The beads were set aside in room temperature until the probes were ready to be pulled down. To collect rRNA, 2 to 3{\textmu}g total RNA and 1 {\textmu}l of the diluted biotinylated probe mix were combined on ice into a final annealing reaction mixture of 1xSSC and 500 {\textmu}M EDTA. All the appropriate volumes were computed using the Excel-based calculator. The RNA and probe mixture was incubated at 70{\textdegree}C for 5 min, and slowly cooled to 25{\textdegree}C at a rate of 1{\textdegree}C per 30 s. The annealed mixture was then added to 30 {\textmu}l of beads that were resuspended in 2× B{\&}W buffer. The mixture was mixed by pipetting and vortexing at medium speed, and followed by incubating for 5 min at room temperature. The reaction mixtures were then vortexed, and incubated at 50{\textdegree}C for 5 min. To pull down the biotinylated probes, the reaction mixtures were placed immediately placed on the magnetic rack. The supernatant was carefully pipetted, placed on ice, and diluted to 200 {\textmu}l in DEPC-H2O. The RNA was purified through ethanol precipitation with 20 {\textmu}l of 3 M NaOAc, 2 {\textmu}l GlycoBlue (Invitrogen), and 600 {\textmu}l ice-cold ethanol  at -20{\textdegree}C for at least 1 h. To pellet RNA, the samples were centrifuged at 4{\textdegree}C for 30 min at 21,000 × g. The pellets were washed twice with 500 {\textmu}l of ice-cold 70\% ethanol, followed by centrifugation at 4{\textdegree}C for 5 min. RNA pellets were then air dried and resuspended in 10 {\textmu}l DEPC-H2O. The yield and rRNA depletion effectiveness was verified with NanoDrop spectrophotometer, and by running 50 ng of total RNA on a Novex 6\% Tris-buffered EDTA (TBE)-urea polyacrylamide gel (Invitrogen). The yield and integrity of the library was checked by running the samples in qPCR using NEBNext Library Quant Kit for Illumina(NEB) and the Bioanalyzer. 

\subsubsection{Library prep and sequencing} 

The RNA library was prepared with NEBNext® Multiplex Oligos for Illumina(NEB) and NEBNext ultra II RNA Library Prep Kit for Illumina(NEB). For the library prep protocol, we followed section 4 of the kit's provided protocol: Protocols for use with Purified mRNA or rRNA Depleted RNA. The quality of the final library was verified by running the samples on high sensitivity Bioanalyzer chip. The samples were pooled to a final concentration of 8.5nM, and were sequenced with NextSeq 150 cycle kit.

\subsection{Methods: Computation of message number}

\label{sec:smmessagenum}

To estimate the message number for gene $i$, defined as the total number of mRNA molecules transcribed per cell cycle,  from the RNA-Seq data, we use the approach we described earlier \cite{Lo2024}. Let the relative number of reads for gene $i$ be $r_i$:
\begin{equation}
r_i = \textstyle\frac{N_i}{N_{\rm tot}},
\end{equation}
where $N_i$ is the reads per kilobase (rpk) for gene $i$ and $N_{\rm tot}$ is the rpk for all genes. We apply two different re-scalings: First we re-scale the relative message abundance to reflect the cellular abundance of the message, and then we scale this number by the ratio of cell cycle duration to mRNA lifetime to estimate the number of times a gene is transcribed per cell cycle. 
For \textit{A.~baylyi}, we use the same scaling factor as \textit{E.~coli}:
\begin{equation}
\mu_{m,i} =  9.4\times 10^4 \cdot r_i,
\end{equation}
where $\mu_{m,i}$ is the estimated message number (number of mRNA molecules transcribed per cell cycle).

To check the consistency of this estimate, we generated histograms for message number for essential and non-essential genes, and compared them to the histograms for \textit{E.~coli}. We expect the distribution of essential message numbers to abut 1 message per cell cycle, while non-essential genes can be expressed at significantly lower levels. The observed distribution are consistent with this expectation. (See Fig.~\ref{fig:onemessage}.) 

\subsection{Results: Comparison of \textit{A.~baylyi} and \textit{E.~coli} gene expression }

\label{sec:smEcvsAb}

Knockout-depletion experiments are not tractable in \textit{E.~coli} and many other model systems. It is therefore difficult to directly test the overabundance hypothesis in these other systems. However, it is possible to determine if \textit{E.~coli} expression patterns are consistent with overabundance. 

If overabundance were specific to \textit{A.~baylyi}, we would expect to see higher relative transcription of lower abundance essential genes in  \textit{A.~baylyi}, where overabundance is  large, relative to \textit{E.~coli} if its expression levels were sufficient. Fig.~\ref{fig:EcolivsAbaylyi} compares the message number between homologues in the two organisms and between growth conditions within a particular organism for all genes.

\section{Robustness Load Trade-Off (RLTO) Model }

\label{sec:RLTO} 

We have provided a detailed description of the Robustness Load Trade-Off (RLTO) Model in Ref.~\cite{Lo2024}; however, in the interest of making this paper self-contained, we provide a concise summary of key elements and results from that paper in this supplementary section.

\subsection{Methods: Detailed description of the noise model} 
\label{sec:noise_intro}

\subsubsection{Stochastic kinetic model for the central dogma.}  

The {canonical steady-state noise model} for the central dogma describes multiple steps in the gene expression process \cite{Paulsson:2000xi,Friedman:2006oh,Taniguchi2010}:
 Transcription generates mRNA messages.  These messages are then translated to synthesize the protein gene products \cite{Crick:1970vy}.  Both mRNA and protein are subject to degradation and dilution \cite{Hargrove:1989vo}. 
 At the single cell level, each of these processes are stochastic. We will model these processes with the stochastic kinetic scheme   \cite{Crick:1970vy}:
\begin{equation}
  \begin{CD}
   {\rm DNA} @>\beta_m>> {\rm mRNA} & @>\beta_p>>  {\rm Protein} \\
     & & @V\gamma_mVV & @V\gamma_pVV\\
     & & \varnothing && &  \; \varnothing, \label{stochmodel}
  \end{CD}
\end{equation}
where $\beta_m$ is the transcription rate (s$^{-1}$), $\beta_p$ is the translation rate (s$^{-1}$), $\gamma_m$ is the message  degradation rate (s$^{-1}$), and $\gamma_p$ is the  protein effective degradation rate (s$^{-1}$).  The message lifetime is $T_m\equiv \gamma_m^{-1}$. For most proteins in the context of rapid growth, dilution is the dominant mechanism of protein depletion and therefore $\gamma_p$ is approximately the growth rate  \cite{KOCH:1955oa,Martin-Perez:2017jx,Taniguchi2010}: $\gamma_p = T^{-1}\ln 2$,  where $T$ is the doubling time. 

\label{sectelegraph}

\subsubsection{Statistical model for  protein abundance.} 

\label{sec:statnoiseS}

Consistent with previous reports \cite{Paulsson:2000xi,Friedman:2006oh}, we find that the distribution of protein number per cell (at cell birth) was described by a gamma distribution \cite{Lo2024}:
\begin{equation}
N_p \sim \Gamma(a,\theta), 
\end{equation}
where $N_p$ is the protein number at cell birth and $\Gamma$ is the gamma distribution, which is parameterized by a scale parameter $\theta$ and a shape parameter $a$. (See Sec.~\ref{sec:gammaconv}.) We refer to this distribution as the \textit{canonical steady-state noise model};
The relation between the four kinetic parameters and these two statistical parameters has already been reported, and have clear biological interpretations \cite{Friedman:2006oh}: The scale parameter: 
\begin{eqnarray} 
\theta = \varepsilon \ln 2, \label{eqn:eff_rates} 
\end{eqnarray}
is proportional to the translation efficiency:
\begin{eqnarray}
\varepsilon \equiv \textstyle \frac{\beta_p}{\gamma_m}, \label{eqn:TE}
\end{eqnarray} 
where  $\beta_p$ is the translation rate and $\gamma_m$ is the message degradation rate.   $\varepsilon$ is understood as the mean number of proteins translated from each message transcribed. 
The shape parameter $a$ can also be expressed in terms of the kinetic parameters \cite{Friedman:2006oh}: 
\begin{equation}
a = \textstyle\frac{\beta_m}{\gamma_p }; \label{eqn:shape2}
\end{equation}
however, we will find it more convenient to express the scale parameter in terms of the cell-cycle message number:
\begin{eqnarray}
    \mu_{m} \equiv \beta_m T =  a \ln 2,
\end{eqnarray}
which can be interpreted as the mean  number of messages transcribed per cell cycle. Forthwith, we will abbreviate this quantity \textit{message number} in the interest of brevity.

\subsection{Methods: Summary of the RLTO model fitness model} 

\subsubsection{Metabolic load in the RLTO model}

To produce a minimal model to study the trade-off between robustness and metabolic load, we must consider both the metabolic cost of transcription and translation. We will write that the metabolic load (in protein equivalents) associated with gene $i$ is:
\begin{equation}
\delta N_i = \lambda \mu_{m,i}+ \mu_{p,i},
\end{equation}  
where $\lambda$ is the message cost, the  metabolic load associated with an mRNA molecule relative to a single protein molecule of the gene product.   
\begin{equation}
\ln \textstyle \frac{k}{k_0} = -\textstyle \frac{(\lambda + \varepsilon_i)\mu_{m,i}}{N_0}. \label{eqn:almost2}
\end{equation}
This equation has an intuitive interpretation: growth slows in proportion to the relative added metabolic load. In resource allocation models \cite{Scott:2010ec}, the capacity of the cell for growth can increase as protein sectors increase in size. In our context, this does not occur since we consider the uncoordinated changes in the levels of single proteins. \textit{I.e.}\ we assume some other protein of factor is rate limiting. See the detailed discussion in Ref.~\cite{Lo2024}.

\label{sec:derivationgr}

\subsubsection{Growth rate with stochastic arrest}
As discussed in Ref.~\cite{Lo2024}, we idealize the slow growth associated with essential proteins falling below threshold as growth arrest. This arrest model has phenomenology consistent with more detailed and realistic models where cells experience a significant growth slowdown rather than true growth arrest \cite{Lo2024}.

In the idealized growth arrest model, if all essential proteins are above threshold, the cell cycle duration $\tau$ is determined by the metabolic load predictions (Eq.~\ref{eqn:almost2}); however, if any essential protein is below threshold, the cell cycle duration is infinite. The probability mass function for the cycle-cycle duration $T$ interpreted as a random variable is therefore:
\begin{equation}
p_T(t) = \begin{cases} P_+, & t = \tau \\
(1-P_+), & t \rightarrow \infty 
\end{cases}.
\end{equation}
As we show in Ref.~\cite{Lo2024}, the growth rate can be computed exactly:
\begin{equation}
k = \tau^{-1} \ln (2P_+).    \label{eqn:dergrowth} 
\end{equation}
As expected, the growth rate goes down as the probability of growth $P_+$ decreases, stopping completely at $P_+=\frac{1}{2}$.
We can then compute the ratio of the growth with ($k$) and without arrest ($k_0$):
\begin{equation}
\textstyle\frac{k}{k_0} = 1+\textstyle\frac{1}{\ln 2}\ln P_+,    \label{eqn:dergrowth2} 
\end{equation}
where $k_0$ is computed by evaluating Eq.~\ref{eqn:dergrowth} at $P_+=1.$

\subsubsection{RLTO  growth rate}

In the RLTO model, we will assume the probability of growth is the probability that all essential protein numbers are above threshold.  We will further assume that each protein number is independent, and therefore: 
\begin{equation}
P_+ = \prod_{i\in {\cal E}} {\rm Pr}\{ N_{p,i}>n_{p,i}\},
\end{equation}
where ${\cal E}$ is the set of essential genes. Clearly, this assumption of independence fails in the context of polycistronic messages. We will discuss the significance of this feature of bacterial cells  elsewhere, but we will ignore it in the current context. As we will discuss, the probability of arrest of any protein $i$ to be above threshold is extremely small. It is therefore convenient to work in terms of the CDFs, which are very close to zero:
\begin{eqnarray}
\ln P_+ &\approx& -\sum_{i\in {\cal E}} \textstyle \gamma^-( \frac{\mu_{m,i}}{\ln 2},\frac{n_{p,i}}{\varepsilon_i\ln 2}), \label{eqn:gammader}
\end{eqnarray}
where $\gamma^-$ is the regularized lower incomplete gamma function (Eq.~\ref{eqn:lower}) and represents the probability of arrest.

\subsubsection{Single-gene equation}
By summing the fitness losses from the metabolic load and cell arrest (Eqs.~\ref{eqn:almost2}, \ref{eqn:dergrowth2},  and \ref{eqn:gammader}), we can write an expression for the growth rate including contributions from essential gene $i$:
\begin{eqnarray}
\ln \textstyle \frac{k}{k_0} &=& -\textstyle\frac{\lambda+ \varepsilon_i}{N_0} \mu_{m,i}   -\textstyle \frac{1}{\ln 2} \gamma^-( \frac{\mu_{m,i}}{\ln 2},\frac{n_{p,i}}{\varepsilon\ln 2}), \label{eq:paw1}
\end{eqnarray}
where the first term on the RHS represents the fitness loss due to the metabolic load and the second term represents the fitness loss due to stochastic cell arrest due to protein $i$ falling below threshold.

\subsubsection{Optimization of transcription for bacteria}
\label{sec:opti}

The  growth rate is: 
\begin{equation}
\ln \textstyle \frac{k}{k_0} =  \textstyle - (\Lambda+\frac{\varepsilon}{N_0})\mu_m - \frac{1}{\ln 2}\gamma^-(\frac{\mu_m}{\ln 2},\frac{n_p}{\varepsilon \ln 2}), \label{SIeqn:growthrate2}
\end{equation}
where $\gamma^-$ is the regularized lower incomplete gamma function (Eq.~\ref{eqn:lower}), which is the CDF of the gamma distribution and represents the probability of arrest due to gene $i$. For bacteria, we consider the special case of optimizing the message number only at fixed translation efficiency \cite{Taniguchi2010,Lo2024}. To determine the optimal transcription level, we set the partial derivative of Eq.~\ref{SIeqn:growthrate2} with respect to $\mu_m$ to zero. The optimum message number $\hat{\mu}_m$ satisfies the equation:
\begin{equation}
\textstyle\frac{(\lambda + \varepsilon) \ln 2}{N_0} =  - [\partial_{\hat{\mu}_m}\gamma( \hat{\mu}_m,\hat{n}_m)]_{\hat{n}_m = \frac{\hat{\mu}_m}{\hat{o}}}.
\end{equation}
We define the relative load:
\begin{equation}
\Lambda \equiv \textstyle\frac{(\lambda + \varepsilon)}{N_0}, \label{eqn:modload}
\end{equation}
and substitute this into the optimum message number equation:
\begin{equation}
\Lambda \ln 2 =  - [\partial_{\hat{\mu}_m}\gamma( \hat{\mu}_m,\hat{n}_m)]_{\hat{n}_m = \frac{\hat{\mu}_m}{\hat{o}}},
\end{equation}
which is solved numerically.

\begin{figure*}
  \centering
    \includegraphics[width=0.95\textwidth]{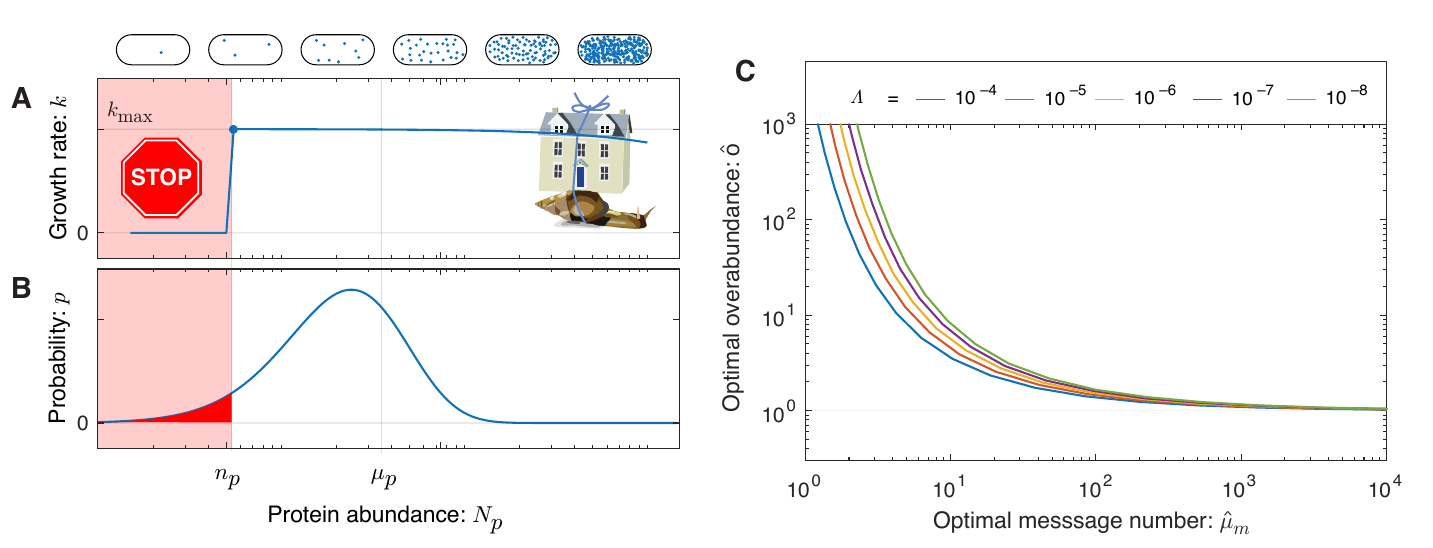}
        \caption{
        \textbf{Panel A: The fitness landscape is asymmetric in the RLTO model.} Motivated by single-cell growth data, cell fitness is modeled using the Robustness-Load Trade-Off model (RLTO).
        In the model, there is a metabolic cost of protein expression which favors low expression; however, growth arrests for protein number $N_p$ smaller than the threshold level $n_p$ (red).  The relative metabolic cost of overabundance is small relative to the cost of growth arrest due to the large number of proteins synthesized, resulting in a highly asymmetric fitness landscape \cite{Lo2024}.  
        \textbf{Panel B: The gene expression process is stochastic.} There is significant cell-to-cell variation in protein abundance ($N_p$) around the mean level ($\mu_p$).  Even for mean expression levels significantly above the threshold level $n_p$, some cells fall below threshold (red). The distribution in protein number is modeled using a gamma distribution \cite{Taniguchi2010}.
        \textbf{Panel C: Overabundance is predicted to optimize cell fitness.}  The asymmetry of the fitness landscape drives the optimal protein expression level to be overabundant ($\mu_p> n_p$). The RLTO model makes a quantitative prediction of the optimal overabundance ($\hat{o}\equiv \hat{\mu}_p/n_p$) as a function of the message number $\mu_m$ and a global parameter, the relative load $\Lambda \approx 10^{-5}$ (red curve).
        Overabundance is predicted to be extremely high ($o \gg 1$) for low expression genes ($\mu_m\approx 1$) and much closer to sufficiency ($o\approx 1$) for high expression genes ($\mu_m\gg 10$).  Although the optimal overabundance depends on the relative load $\Lambda$, its qualitative dependence is unchanged over orders of magnitude in variation of the parameter. } \label{fig:RLTO}
\end{figure*}

\subsubsection{Estimate of the  relative load in bacterial cells}

In bacterial cells, we will assume a constant translation efficiency model. We therefore use the modified relative load formula (Eq.~\ref{eqn:modload}) to estimate $\Lambda$. We will assume that the load is dominated by proteins and messages:  
\begin{eqnarray}
N_0  &=& \sum_i (\lambda + \varepsilon) {\mu_{m,i}} = (\lambda + \varepsilon) N_m,
\end{eqnarray}
where $N_m$ is the total number of messages. We can then solve this equation for  $\Lambda$:
\begin{equation}
\hat{\Lambda}  = \textstyle\frac{\lambda + \varepsilon}{N_0} = \frac{1}{N_m} \approx 10^{-5},
\end{equation}
based on the total message number estimate for \textit{E.~coli} \cite{Lo2024}.

\subsection{Results: The fitness landscape of the RLTO model is highly asymmetric}

In the RLTO model, the fitness landscape for a single cell is determined by an asymmetric fitness landscape: protein underabundance is extremely costly due to the risk of growth arrest, while the cost of protein overabundance is only associated with an increase in metabolic load. (See Fig.~\ref{fig:RLTO}A.) 
Na\"ively, this tradeoff predicts that the cell maximizes its fitness by simply expressing just above the minimum protein threshold for function \cite{Belliveau:2021pj}. However, achieving growth robustness at a population level is nontrivial. Gene expression is stochastic \cite{Raser:2005we}, leading to significant cell-to-cell variation in protein numbers, which we model with a gamma distribution (Fig.~\ref{fig:RLTO}B) \cite{Friedman:2006oh,Paulsson:2000xi}. Therefore, the strong asymmetry of the fitness landscape predicts protein overabundance. 

\subsection{Results: The RLTO model predicts overabundance is optimal for low-expression proteins}

The optimal regulatory program for transcription and translation ($\mu_m$ and $\varepsilon$ values) can be predicted analytically. The values are determined by a single global parameter, the relative load $\Lambda$, and the gene-specific threshold number $n_p$. The threshold number is not directly observable experimentally; instead we predict the optimal overabundance $o$, defined as the ratio of the mean protein number to the threshold number: 
\begin{equation}
o \equiv \textstyle \mu_p/n_p.
\end{equation}

As shown in \cite{Lo2024}, by taking partial derivatives of the relative growth rate (Eq.~\ref{SIeqn:growthrate2}) with respect to message number and translation efficiency, respectively, we can define the optimal overabundance:
\begin{equation}
\hat{o} \equiv \textstyle\frac{\hat{\mu}_p}{n_p} = \frac{\hat{\varepsilon} \hat{\mu}_m}{n_p},
\end{equation}
in the large multiplicity limit where the overall metabolic load is much smaller than the metabolic load for a single gene: $N_0 \gg (\lambda+\hat{\varepsilon})\hat{\mu}_m$. The optimal overabundance can be rewritten to find the optimization condition for message number:
\begin{eqnarray}
\Lambda \ln 2  &=& \textstyle -\partial_{\hat{\mu}_m} \gamma( \frac{\hat{\mu}_m}{\ln 2},\frac{\hat{\mu}_m}{\hat{o}\ln 2}). \label{eqn:oeqn}
\end{eqnarray}

As seen in Fig.~\ref{fig:RLTO}C, the RLTO model generically predicts that for a range of relative loads, the optimal protein fraction is overabundant ($o>1$); however, overabundance is not uniform for all proteins, but rather depends on transcription. For highly-transcribed genes ($\mu_m\gg 1$), the overabundance is predicted to be quite small ($o\approx 1$); however, for lowly-transcribed genes (message numbers approaching unity), the overabundance is predicted to be extremely high $(o\gg 1)$.

\subsection{Discussion: Does the detailed form of the fitness landscape affect RLTO predictions?} 
\label{sec:detailedform}
It is important to emphasize that the detailed mathematical form of the RLTO model is not essential to generate the predicted phenomenology. For instance, changes in the functional form of the protein expression noise, the metabolic load, or a more realistic model of the fitness landscape do not significantly change the predictions of the model. It is the strong asymmetry of the fitness landscape that is required to  predict protein overabundance \cite{Lo2024}.

\section{Methods: Statistical procedures}

In this section, we provide a summary of statistical approaches that are common to the analyses in the paper.

\subsection{Maximum Likelihood Estimation } \label{Method:Unc}
The maximum likelihood (\textit{i.e.}~minimum information) estimates (MLE) of the parameters are defined:
\begin{equation}
\hat{\theta}^i = \arg \min_{\theta^i} h({\rm data}|\theta^i). \label{eqn:MLE}
\end{equation}
In all instances, these optimizations are performed numerically, either by direct minimization of the Shannon information ($h$), or for normal models, by least-squares minimization.

\subsection{Parametric uncertainty estimates} 

To estimate the parameter uncertainty in the analysis of datasets,  we use the Cramer-Rao bound to estimate of the uncertainty from the Fisher information \cite{CoxHink74}:
\begin{equation}
\sigma_{\theta^i} = \sqrt{[\hat{I}^{-1}]^{ii}}, \label{eqn:uncertainty}
\end{equation}
where $\sigma_{\theta^i}$ is the estimate of the standard error for parameter $\theta^i$, $\hat{I}$ is the estimator of the Fisher information, and $[\hat{I}^{-1}]^{ii}$ is the $ii$ component of the inverse Fisher information matrix. For each statistical model, we describe how the Fisher information is estimated in detail (Hessian or Jacobian \textit{etc}).

\subsection{ Null-hypothesis-testing approach}

For null-hypothesis testing, we define two sequential null-hypothesis tests of nested statistical models. If the initial null hypothesis is rejected, we then interpret the initial alternative hypothesis as the updated null hypothesis and adopt the remaining model as the alternative hypothesis. For each test, we will use a Likelihood Ratio Test (LRT) where we define the test statistic $\lambda$ in terms of the Shannon information:
\begin{equation}
\lambda = h_0-h_1, \label{eqn:teststat}
\end{equation}
where $h_0$ and $h_1$ are the Shannon information for the null and alternative hypotheses respectively. We will assume the Wilks' theorem: \textit{I.e.} the test statistic $\Lambda$ under the null hypothesis will have a chi-squared distribution \cite{Wilks1938,CaseBerg:01}:
\begin{equation}
\Lambda \sim \textstyle \frac{1}{2} \chi^2_{\Delta K},
\end{equation}
where the degrees-of-freedom $\Delta K = 1$ is equal to the difference between the dimension of the alternative and null models. (The factor of 1/2 appears in this equation, since the test statistic is defined by the Shannon information difference rather than the deviance \cite{CoxHink74}.) The p-value can then be computed: 
\begin{equation}
p = {\rm Pr} \{ \Lambda > \lambda \} = \gamma^+(\textstyle\frac{1}{2}\Delta K, \lambda ), \label{eqn:pval}
\end{equation}
where $\gamma^+$ is the upper regularized incomplete gamma function (Eq.~\ref{eqn:survchi}),  $\Delta K = 1$ is the difference in model dimensions, and $\lambda$ is the test statistic \cite{CoxHink74}.

\section{Distributions and conventions}

\subsection{Gamma distribution conventions}
\label{sec:gammaconv}
There are a number of  conflicting conventions for the gamma function and distribution arguments. We will use those defined on Wikipedia and the CRC Encyclopedia of Mathematics \cite{Weisstein2009}. 

The gamma distributed random variable $X$ will be written:
\begin{equation}
    X\sim \Gamma( a,\theta),
\end{equation}
where $a$ is the shape parameter and $\theta$ is the scale parameter. The PDF of the distribution is:
\begin{equation}
p_X(x|a,\theta) \equiv \textstyle \frac{x^{a-1}}{\theta^{a}\Gamma(a)} e^{-x/\theta}, \label{eqn:pdfgamma}
\end{equation}
where $\Gamma(a)$ is the gamma function. The CDF is therefore:
\begin{eqnarray}
P_X(x|a,\theta) &\equiv& {\rm Pr}\{ X< x|a,\theta\},\\
&=& \int_0^x {\rm d}x'\ p_\Gamma(x'|a,\theta),\\ 
&=& \int_0^{x/\theta} {\rm d}x''\ \textstyle \frac{x''^{a-1}}{\Gamma(a)} e^{-x},\\
&=& \gamma^-( a, x/\theta), \label{eqn:lower}
\end{eqnarray}
where $\gamma^-$ is the regularized lower incomplete gamma function. 
The survival function is:
\begin{eqnarray}
{\rm Pr}\{ X> x|a,\theta\} &=& 1-P_X(x|a,\theta), \\
&=& \int_{x/\theta}^\infty {\rm d}x''\ \textstyle \frac{x''^{a-1}}{\Gamma(a)} e^{-x},\\
&=& \gamma^+( a, x/\theta), 
\end{eqnarray}
where $\gamma^+$ is the regularized upper incomplete gamma function.

\subsection{Chi-squared distribution conventions}

In statistical null hypothesis testing, the chi-squared distribution arises in the context of the Likelihood Ratio Test (LRT).
Let $Y$ be distributed like a chi-squared with $k$ degrees of freedom:
\begin{equation}
Y \sim \chi^2_k,    
\end{equation}
where the PDF is: 
\begin{equation}
p_Y(y|k) = \textstyle \frac{1}{2^{k/2}\Gamma(k/2)} y^{k/2-1}e^{-y/2},   
\end{equation}
where $\Gamma$ is the gamma function. The CDF is therefore:
\begin{eqnarray}
P_Y(y|k) &\equiv& {\rm Pr}\{ Y< y|k\},\\
&=& \int_0^y {\rm d}y'\ p_Y(y'|k),\\ 
&=& \int_0^{x} {\rm d}x'\ p_X(x'|\textstyle\frac{k}{2},2),\\
&=& \gamma^-(  \textstyle\frac{k}{2}, \textstyle\frac{y}{2}),
\end{eqnarray}
where $p_X$ is the PDF of the gamma distribution (Eq.~\ref{eqn:pdfgamma}) and $\gamma^-$ is the regularized lower incomplete gamma function. 
The survival function is:
\begin{eqnarray}
{\rm Pr}\{ Y> y|k\} &=& \gamma^+( \textstyle\frac{k}{2}, \textstyle\frac{y}{2}), \label{eqn:survchi}
\end{eqnarray}
where $\gamma^+$ is the regularized upper incomplete gamma function.

\section{Description of supplementary data}
\subsection{Data Tables}

\idea{Data S1}: Overabundance for all genes as measured by TFNseq analysis. The original TFNseq data was previously generated by the Manoil lab \cite{Gallagher:2020sp}. Format: Open Document Format (ODS). 

\idea{Data S2}: A list of essential genes ranked by overabundance. Format: Open Document Format (ODS). 

\idea{Data S3}: Representative single-cell imaging-based cell cytometry data for wild-type \textit{A.~baylyi} proliferating on minimal media (Km\textsuperscript{-}) from a single progenitor cell (Sec.~\ref{sec:wt}). Format: Open Document Format (ODS). 

\idea{Data S4}: Representative single-cell imaging-based cell cytometry data for \textit{A.~baylyi} $\Delta$\textit{IS}  proliferating on minimal media (Km\textsuperscript{+}) from a single progenitor cell in a knockout-depletion experiment (Sec.~\ref{sec:IS}). Format: Open Document Format (ODS). 

\idea{Data S5}: Representative single-cell imaging-based cell cytometry data for \textit{A.~baylyi}  $\Delta$\textit{dnaA} proliferating on minimal media (Km\textsuperscript{+}) from a single progenitor cell in a knockout-depletion experiment (Sec.~\ref{sec:dnaA}). Format: Open Document Format (ODS). 

\idea{Data S6}: Representative single-cell imaging-based cell cytometry data for \textit{A.~baylyi} $\Delta$\textit{dnaN}  proliferating on minimal media (Km\textsuperscript{+}) from a single progenitor cell in a knockout-depletion experiment (Sec.~\ref{sec:dnaN}). Format: Open Document Format (ODS). 

\idea{Data S7}: Representative single-cell imaging-based cell cytometry data for \textit{A.~baylyi}  $\Delta$\textit{murA} proliferating on minimal media (Km\textsuperscript{+}) from a single progenitor cell in a knockout-depletion experiment (Sec.~\ref{sec:murA}). Format: Open Document Format (ODS). 

\idea{Data S8}: Representative single-cell imaging-based cell cytometry data for \textit{A.~baylyi} $\Delta$\textit{ftsN} proliferating on minimal media (Km\textsuperscript{+}) from a single progenitor cell in a knockout-depletion experiment (Sec.~\ref{sec:ftsN}). Format: Open Document Format (ODS). 

\subsection{Annotated sequences}

\idea{Data S9}: The annotated sequence of the DnaN fluorescent fusion \textit{YPet-dnaN}.  Format: Genbank file.

\subsection{Supplemental movies}

\idea{Movie S1}: Wild-type \textit{A.~baylyi} proliferating on minimal media (Km\textsuperscript{-}). Frame rate: 1 frame/2 min. (Sec.~\ref{sec:wt}.) Raw images.
\href{https://youtu.be/TTbY-Ry4Xho}{Youtube.}

\idea{Movie S2}: Wild-type \textit{A.~baylyi} proliferating on minimal media (Km\textsuperscript{-}). Frame rate: 1 frame/2 min. (Sec.~\ref{sec:wt}.) Annotated/segmented images.
\href{https://youtu.be/9Ak6LL3lS9I}{Youtube.}

\idea{Movie S3}:  \textit{A.~baylyi} $\Delta$\textit{IS} proliferating on minimal media (Km\textsuperscript{+}) in a knockout-depletion experiment. Frame rate: 1 frame/3 min. (Sec.~\ref{sec:IS}.) Raw images.
\href{https://youtu.be/flyIQYcxzTk}{Youtube.}

\idea{Movie S4}: \textit{A.~baylyi} $\Delta$\textit{IS}  proliferating on minimal media (Km\textsuperscript{+}) in a knockout-depletion experiment. Frame rate: 1 frame/3 min. (Sec.~\ref{sec:IS}.) Annotated/segmented images.
\href{https://youtu.be/FgIF9LsCVmA}{Youtube.}

\idea{Movie S5}: \textit{A.~baylyi}  $\Delta$\textit{dnaA} proliferating on minimal media (Km\textsuperscript{+}) in a knockout-depletion experiment. Frame rate: 1 frame/2 min. (Sec.~\ref{sec:dnaA}.) Raw images.
\href{https://youtu.be/AfkDaYZ2lII}{Youtube.}

\idea{Movie S6}: \textit{A.~baylyi} $\Delta$\textit{dnaA} proliferating on minimal media (Km\textsuperscript{+}) in a knockout-depletion experiment. Frame rate: 1 frame/2 min. (Sec.~\ref{sec:dnaA}.) Annotated/segmented images. 
\href{https://youtu.be/EjYRzKXhfe0}{Youtube.}

\idea{Movie S7}: \textit{A.~baylyi} $\Delta$\textit{dnaN}  proliferating on minimal media (Km\textsuperscript{+}) in a knockout-depletion experiment. Frame rate: 1 frame/9 min. (Sec.~\ref{sec:dnaN}.) Raw images.
\href{https://youtu.be/h_1zW7Pntvw}{Youtube.}

\idea{Movie S8}:  \textit{A.~baylyi} $\Delta$\textit{dnaN} proliferating on minimal media (Km\textsuperscript{+}) in a knockout-depletion experiment. Frame rate: 1 frame/9 min. (Sec.~\ref{sec:dnaN}.) Annotated/segmented images.
\href{https://youtu.be/i58BSBwx_hw}{Youtube.}

\idea{Movie S9}:  \textit{A.~baylyi} $\Delta$\textit{murA} proliferating on minimal media (Km\textsuperscript{+}) in a knockout-depletion experiment. Frame rate: 1 frame/2 min. (Sec.~\ref{sec:murA}.) Raw images. \href{https://youtu.be/Jb6lzQ1Nsp8}{Youtube.}

\idea{Movie S10}: \textit{A.~baylyi} $\Delta$\textit{murA}  proliferating on minimal media (Km\textsuperscript{+}) in a knockout-depletion experiment. Frame rate: 1 frame/2 min. (Sec.~\ref{sec:murA}.) Annotated/segmented images. \href{https://youtu.be/E7f_Pi1inkw}{Youtube.}

\idea{Movie S11}:  \textit{A.~baylyi}  $\Delta$\textit{ftsN} proliferating on minimal media (Km\textsuperscript{+}) in a knockout-depletion experiment. Frame rate: 1 frame/2 min. (Sec.~\ref{sec:ftsN}.) Raw images.
\href{https://youtu.be/XnqjFlfm330}{Youtube.}

\idea{Movie S12}:  \textit{A.~baylyi} $\Delta$\textit{ftsN} proliferating on minimal media (Km\textsuperscript{+}) in a knockout-depletion experiment. Frame rate: 1 frame/2 min. (Sec.~\ref{sec:ftsN}.) Annotated/segmented images.
\href{https://youtu.be/HFGs1jlVJ7c}{Youtube.}


\end{document}